\documentclass[twoside,leqno,twocolumn]{article}

\usepackage[letterpaper]{geometry}
\makeatletter
\ifx\XeTeXrevision\@undefined\else
  \expandafter\def\csname ver@hypdvips.sty\endcsname{2019/07/31 v3.03}
  \expandafter\def\csname ver@breakurl.sty\endcsname{2013/04/10 v1.40}
\fi
\makeatother
\usepackage{siamproceedings}

\usepackage[T1]{fontenc}
\usepackage{amsmath,amssymb,amsfonts}
\usepackage{graphicx}
\graphicspath{{./}}
\usepackage{booktabs}
\usepackage{multirow}
\usepackage[noend]{algpseudocode}
\usepackage{tikz}
\usetikzlibrary{arrows.meta}
\usepackage{xcolor}
\usepackage{comment}
\usepackage{xspace}
\usepackage{numprint}
\usepackage{needspace}

\newsiamthm{conjecture}{Conjecture}
\newsiamremark{remark}{Remark}
\crefname{algorithm}{Algorithm}{Algorithms}
\Crefname{algorithm}{Algorithm}{Algorithms}

\newcommand\relatedversion{}

\newif\ifblind
\blindfalse
\newcommand{\TheTitle}{A Parallel Greedy Algorithm for Maximal $b$-Matching in Hypergraphs}
\ifblind
  \newcommand{\TheAuthors}{Anonymous}
  \title{\Large \TheTitle\relatedversion}
  \author{}
  \date{}
\else
  \newcommand{\TheAuthors}{Saeed Bakhshan, Daniel Grosu, Henrik Reinst\"adtler, and Christian Schulz}
  \title{\Large \TheTitle\relatedversion\thanks{This work was supported by the Deutsche Forschungsgemeinschaft (DFG, German Research Foundation) -- project number 543787284.}}
  \author{Saeed Bakhshan\thanks{Wayne State University, Detroit, MI, USA.}
    \and Daniel Grosu\thanks{Wayne State University, Detroit, MI, USA.}
    \and Henrik Reinst\"adtler\thanks{Heidelberg University, Heidelberg, Germany.}
    \and Christian Schulz\thanks{Heidelberg University, Heidelberg, Germany.}
  }
  \date{}
\fi

\algrenewcommand\algorithmicrequire{\textbf{Input:}}
\algrenewcommand\algorithmicensure{\textbf{Output:}}
\algnewcommand\algorithmicinput{\textbf{Input:}}
\algnewcommand\INPUT{\item[\algorithmicinput]}
\algnewcommand\algorithmicoutput{\textbf{Output:}}
\algnewcommand\OUTPUT{\item[\algorithmicoutput]}

\algnewcommand\algorithmicparfor{\textbf{parfor}}
\algnewcommand\algorithmicpardo{\textbf{do}}
\algnewcommand\algorithmicendparfor{\textbf{end\ parfor}}
\algdef{SE}[PARFOR]{ParFor}{EndParFor}[1]%
  {\algorithmicparfor\ #1\ \algorithmicpardo}%
  {\algorithmicendparfor}
\algtext*{EndParFor}

\makeatletter
\algnewcommand\LongState[1]{%
  \State\parbox[t]{\dimexpr\linewidth-\ALG@thistlm\relax}{%
    \parindent\z@ \hangindent\algorithmicindent \hangafter\@ne
    \raggedright #1\strut}}
\makeatother

\newcommand{\code}[1]{\texttt{#1}}
\newcommand{\edges}{\code{edges}}
\newcommand{\topp}{\code{top}}
\newcommand{\nextp}{\code{next}}
\newcommand{\Nv}{\code{N}}
\newcommand{\bcount}{\code{bcount}}
\newcommand{\counter}{\code{counter}}
\newcommand{\done}{\code{done}}
\newcommand{\finished}{\code{finished}}
\newcommand{\Sat}{\code{sat}}
\newcommand{\vote}{\code{vote}}

\newcommand{\true}{\code{true}}

\newcommand{\getRandomPermutation}{\textsc{getRandomPermutation}}
\newcommand{\sortop}{\textsc{sort}}
\newcommand{\increment}{\textsc{increment}}

\newcommand{\findNext}{\textsc{findNext}}

\newcommand{\deleteop}{\textsc{delete}}
 \newcommand{\algoFont}[1]{\textsf{#1}\xspace}
 \newcommand{\dataFont}[1]{\textsc{#1}\xspace}
\newcommand{\seqGreedy}{\mbox{\textsf{SeqGreedy}}\xspace}

\newcommand{\parallelBatchBMatch}{\algoFont{PB-bM}}

 \newcommand{\competitorGREEDYSLIM}{\algoFont{SeqGreedy}}
 \newcommand{\competitorGREEDY}{\algoFont{SeqGreedy}}
 \newcommand{\competitorHLMCUDA}{\algoFont{HLM-CUDA}}

\hypersetup{pdftitle={\TheTitle}, pdfauthor={\TheAuthors}}

\begin{document}

\maketitle

\fancyfoot[R]{\scriptsize{Copyright \textcopyright\ 2027 by SIAM\\
Unauthorized reproduction of this article is prohibited}}

\begin{abstract}
We present \parallelBatchBMatch, a parallel batch-vote algorithm for maximal
$b$-matching in hypergraphs.
The algorithm is deterministic for a given ordering of edges and
produces the same output as the sequential greedy algorithm, i.e., the lexicographically
first maximal $b$-matching. Under a uniformly random priority order it performs
$\mathcal{O}(m')$ expected work in the binary fork--join model, where $m'$ is the
number of pins, the total number of vertex--edge incidences.  
The algorithm performs at most $m$ rounds for every priority order, where $m$ is the number of edges.
Every round of the algorithm has $\mathcal{O}(\log^2 m')$ depth with high probability, and thus the total depth of the algorithm is $\mathcal{O}(m \log^2 m')$ with high probability.
We implement the parallel
algorithm in CUDA and evaluate it on real-world hypergraphs with capacities $b \in
\{1, 3, 5\}$. The parallel algorithm computes the same matchings as the greedy algorithms 
and for the case of $b = 5$ it is up to $19.07$ times faster than the existing state-of-the art greedy algorithm.
\end{abstract}



\smallskip
\noindent\textbf{Code \& experimental data.} We provide the code at \href{https://github.com/HeiHGM/Parallel}{github.com/HeiHGM/Parallel}, an archived version can be found at \href{https://doi.org/10.5281/zenodo.24085030}{10.5281/zenodo.24085030}.
\section{Introduction. }\label{sec:intro}

Given a hypergraph $G = (V, E)$ with $n = |V|$ vertices and $m = |E|$
hyperedges, where each edge $e \in E$ is a set $V(e)$ of vertices, and a
\emph{capacity function} $b : V \to \mathbb{Z}_{\geq 1}$, a \emph{$b$-matching}
is a set $M \subseteq E$ such that no vertex~$v$ is incident to more than~$b(v)$
edges of~$M$. A $b$-matching is \emph{maximal} if no edge in $E \setminus M$ can
be added without exceeding the capacity of one of its vertices. The
\emph{maximal hypergraph $b$-matching} problem asks to compute \hbox{such a maximal
$b$-matching.}

Hypergraph $b$-matchings arise naturally in several application domains. In
scheduling, consider assigning jobs to machines where each job needs a group of
machines at once, and each machine $v$ can run at most $b(v)$ jobs
concurrently. This models settings in which a task holds several resources at
the same time and no resource may be oversubscribed. $b$-matchings are used in
online advertising, where advertiser budgets act as
capacities~\cite{buchbinder2007adauctions,mehta2007adwords}, and in
reconfigurable datacenter networks~\cite{bienkowski2020online}. In machine
learning they build balanced similarity graphs~\cite{jebara2009graph}, enforce
diversity in recommendation~\cite{ahmed2017diverse}, and provide
\hbox{privacy-preserving anonymization~\cite{choromanski2013anonymity}.}

Computing a \emph{maximum} $b$-matching in hypergraphs is NP-hard, and even hard
to approximate~\cite{elouali2012hyperbmatch,parekh2016genhypermatch}. This
motivates the study of \emph{maximal} $b$-matchings, which can be computed in
linear time by a \hbox{simple sequential greedy algorithm (called \seqGreedy).}

\seqGreedy processes the edges in some fixed order and adds
each edge whose vertices are all having some capacity available. It runs in
$\mathcal{O}(m')$ time and produces a maximal $b$-matching, where
$m' = \sum_{e \in E} |V(e)|$ is the number of vertex--edge incidences of $G$. The
particular matching obtained depends on the order in which the edges are
processed. We call such an order a \emph{priority order} $\pi$, and the result
the \emph{lexicographically first} maximal $b$-matching with respect to $\pi$.
While \seqGreedy underlies widely used $b$-matching
heuristics~\cite{grossmann2026hyperbmatch, mestre2006kextendible}, these are all
sequential, and the question of how to parallelize them efficiently has not been
addressed.

When $b(v) = 1$, for all $v\in V$, and every edge is a pair, the problem is the maximal matching in graphs,
for which parallel algorithms have been studied extensively. Israeli and
Itai~\cite{israeli1986fast} gave a randomized maximal matching algorithm that
achieves $\mathcal{O}(\log n)$ expected depth. However, these algorithms do not
preserve the output of \seqGreedy. In many settings a
deterministic output that matches the sequential result is desirable, because it
allows switching between sequential and parallel execution based on available
resources while returning the same matching~\cite{BB}.

A natural question is whether \seqGreedy can
be parallelized efficiently while producing the same output. To the best of our
knowledge, no work-efficient parallel algorithm that computes the
lexicographically first maximal $b$-matching in both graphs and hypergraphs 
has been proposed prior to this work. 
One might consider splitting each vertex $v$ into $b(v)$ copies, which
reduces $b$-matching to matching with unit capacities, and then applying the
algorithm of Blelloch and Brady~\cite{BB}. However, a hyperedge $(u,v)$ with two vertices 
then expands into a complete bipartite graph between the copies of~$u$ and~$v$, inflating the
edge count by $\Theta(\beta^2)$, where $\beta = \max_{v \in V} b(v)$ is the largest
capacity. The blow-up is $\beta^{|V(e)|}$ for a hyperedge~$e$, making the
reduction not work-efficient. The construction also fails to preserve the
lexicographically first order, since a matching may use one original edge
several times, once per matched pair of copies. Furthermore, since computing the
lexicographically first maximal independent set is P-complete for worst-case
orderings~\cite{cook1985taxonomy}, efficient parallelization of the
lexicographically first maximal $b$-matching may also be difficult in the worst
case. However, the situation changes when we consider random orderings.

Under a random order on edges the greedy algorithm for maximal matching in hypergraphs~\cite{BFS,FN} has polylogarithmic \emph{dependence depth}, the number of rounds obtained by repeatedly matching, in parallel, every
edge with no higher-priority conflict, where the priority of edges is given by the 
random order. 
Blelloch and Brady~\cite{BB} turned this into a work-efficient parallel maximal hypergraph
matching algorithm that reproduces the output of \seqGreedy for unit
capacities.
Considering vertex capacities change this dependence structure. A vertex is removed only when it saturates, not when it is
first matched, thus an edge can wait on~$b(v)$ earlier decisions at each of its
vertices instead of one. %

\paragraph{Contributions.}
Building on this framework, we make the following contributions:
\begin{itemize}
  \item We propose \parallelBatchBMatch, a parallel algorithm for maximal
        $b$-matching in hypergraphs in which each vertex votes for all of its
        remaining capacity at once rather than one slot at a time
        (\Cref{sec:batch-vote}).
  \item We prove that it returns the lexicographically first maximal
        $b$-matching for every priority order, no matter how the rounds are
        scheduled.
        and that it performs $\mathcal{O}(m')$ expected work in the
        binary fork--join model 
        (\Cref{sec:batch-vote}).
  \item %
        We show that \parallelBatchBMatch performs at most $m$ rounds for every
        priority order, where $m$ is the number of edges. Every round has
        $\mathcal{O}(\log^2 m')$ depth with high probability (w.h.p.). Thus,
        \parallelBatchBMatch has a depth of $\mathcal{O}(m \log^2 m')$
        w.h.p.\ (\Cref{sec:analysis}).
  \item We implement the algorithm in CUDA and evaluate it on real-world
        hypergraphs with capacities $b \in \{1, 3, 5\}$, where depending on the priority it is up to
        $19.07$ times faster than a sequential greedy implementation
        (\Cref{sec:experiments}), while delivering the same matching.
\end{itemize}

\section{Related work. }\label{sec:related}
We summarize here the results our algorithm builds on and the closest parallel
$b$-matching algorithms.  Appendix~\ref{sec:related-full} provides a more comprehensive 
discussion of the related work.

\paragraph{Greedy matching and dependence depth.}
Reproducing the solution obtained by a sequential greedy algorithm amounts to
computing the lexicographically first maximal solution. For the
lexicographically first maximal independent set this is
P-complete~\cite{cook1985taxonomy,miyano1989lexicographically}, so such problems
are in general resistant to parallelism under a fixed adversarial order. The
standard remedy is to randomize the order. Blelloch, Fineman, and
Shun~\cite{BFS} proved that on \emph{any} graph a uniformly random order yields a
\emph{dependence depth} of $\mathcal{O}(\log^2 m)$
w.h.p., where the dependence depth is the number of rounds obtained by
repeatedly matching, in parallel, every edge having no higher-priority conflict.
Fischer and Noever~\cite{FN} tightened this bound to $\mathcal{O}(\log m)$
w.h.p. Building on this framework, Blelloch and Brady~\cite{BB} obtained a
work-efficient parallel algorithm for maximal matching in hypergraphs with
$\mathcal{O}(m')$ expected work and $\mathcal{O}(\log^2 m)$ depth w.h.p., which
returns exactly the solution of \seqGreedy for unit capacities. Their approach
is the starting point for our algorithm. No analogue of these round bounds is
known once vertex capacities are considered. 
In \Cref{sec:analysis} we give a
simple bound on the number of rounds of our algorithm, and we leave the derivation of a tighter
bound to future work.

\paragraph{Parallel matching and $b$-matching.}
Parallel maximal matching and maximal independent set are classical
problems~\cite{alon1986fast,israeli1986fast,karp1985fast,luby1986simple}, and the
degree-constrained generalization appeared
early~\cite{garrido1992fmatchings}. Each of these algorithms computes \emph{some}
maximal solution, which depends on the schedule. Parallel algorithms for
hypergraph matching have been developed along several
lines~\cite{dhulipala2024parallel,hanguir2021distributed,harris2020distributed,reinstadtler2026hypergraph},
but none of them accounts for vertex capacities. For $b$-matching the parallel
literature offers only approximation guarantees, and it concerns ordinary graphs
almost exclusively. The b-Suitor
family~\cite{khan2016bsuitorsc,khan2016bsuitorsisc} computes
$\tfrac12$-approximations that scale to billions of edges, with a GPU
implementation~\cite{naim2018gpubmatch}. In the massively parallel (MPC) model,
Ghaffari, Grunau, and Mitrovi\'{c}~\cite{ghaffari2022mpcbmatch} gave a
$(1+\varepsilon)$-approximation in $\mathcal{O}(\log\log\bar d)$ rounds, where
$\bar d$ is the average degree.

\begin{table*}[t]
\footnotesize
\caption{Parallel maximal matching and $b$-matching algorithms. 
(Unless noted, the bounds hold under a uniformly random priority order, w.h.p.\ for depth and rounds and in expectation for work. The label ``rounds'' is shown
where no work--depth bound is available for the MPC (Massively Parallel Computing model). 
$\Delta$ is the maximum degree, $r$ is the maximum edge size, $\bar d$ is the average degree, and $B=\sum_v b(v)$.
The $b(v)$-split row is the vertex-splitting reduction to ordinary
matching. $^{\ddagger}$b-Suitor's bounds are in \emph{expectation} and hold under
\emph{uniformly} random edge weights, and are extended to
general $b$ in~\cite{khan2018bedgecover} only by a sketched reduction to the
unit-capacity case.)} 
\label{tab:related}
\begin{center}
\setlength{\tabcolsep}{2.5pt}
\begin{tabular}{@{}lccccc@{}}
\toprule
Algorithm & $b\ge1$ & Hypergraph & Greedy Output & Work & Depth \\
\midrule
Garrido et al.~\cite{garrido1992fmatchings} ($f$-match) & yes & no & no & -- & $\mathcal{O}(\log^2 n)$ \\ %
Khan et al.~\cite{khan2018bedgecover} ($b$-Suitor) & yes & no & no & $\mathcal{O}(\beta B)^{\ddagger}$ & $\mathcal{O}(\log\Delta\log B)^{\ddagger}$ \\
Ghaffari et al.~\cite{ghaffari2022mpcbmatch}      & yes & no  & no  & --              & $\mathcal{O}(\log\log\bar d)$ rounds \\
$b(v)$-split $+$Blelloch and Brady~\cite{BB}                    & yes & no  & no  & $\mathcal{O}(\beta^2 m')$ & $\mathcal{O}(\log^2 m)$ \\
Reinst\"adtler et al.~\cite{reinstadtler2026hypergraph}  & no  & yes & no  & $\mathcal{O}(m'\log m)$   & $\mathcal{O}(\log\log\Delta\log m)$\\ %
Blelloch and Brady~\cite{BB}                         & no  & yes & yes & $\mathcal{O}(m')$         & $\mathcal{O}(\log^2 m)$ \\
\midrule
\parallelBatchBMatch{} ({\bf this paper})          & yes & yes & yes & $\mathcal{O}(m')$         & $\mathcal{O}(m \log^2 m')$ \\
\bottomrule
\end{tabular}
\end{center}

\end{table*}

None of the parallel algorithms discussed above computes the lexicographically-first maximal 
$b$-matching in hypergraphs. This is the gap that our parallel algorithm closes.
Table~\ref{tab:related} lists only the \emph{parallel} algorithms, each covering some subset of the three
properties listed in columns two to four, i.e., vertex capacities ($b\ge1$), hypergraphs, and the
greedy output. Our parallel algorithm covers all three properties and is listed in the last row of the table. 

\section{Preliminaries. }\label{sec:prelim}
\subsection{Hypergraphs and $b$-matching. }

Let $G = (V, E)$ be a hypergraph with $n = |V|$ vertices and $m = |E|$
(hyper)edges. Each edge $e \in E$ is a nonempty set of vertices denoted by
$V(e) \subseteq V$. The vertices in $V(e)$ are called the \emph{endpoints} of $e$. We
extend this notation to any set of edges~$S$, that is,  $V(S) = \bigcup_{e \in S} V(e)$. 
An ordinary graph is the special case of two-vertex
edges. A \emph{capacity function} $b : V \to \mathbb{Z}_{\geq 1}$
gives each vertex $v$ a positive integer capacity $b(v)$, and we write
$\beta = \max_{v \in V} b(v)$~\cite{grossmann2026hyperbmatch}. No vertex can lie
on more edges than it is incident to. We therefore assume
throughout that $b(v)$ is at most the degree of vertex $v$ for all $v\in V$. 
Thus, $\beta$ is bounded by the \hbox{maximum degree in~$G$, and hence by~$m$.}

A \emph{$b$-matching} is a set $M \subseteq E$ in which no vertex~$v$ belongs to
more than~$b(v)$ edges of~$M$. It is \emph{maximal} when every edge outside $M$
has at least one vertex already at its capacity, so that adding the edge would exceed 
the capacity of that vertex. The
case of \emph{unit capacities}, i.e., $b(v) = 1$ for every vertex $v \in V$, 
represents the maximal matching problem in hypergraphs. 
In this paper, we focus on 
the \emph{maximal} $b$-matching problem in hypergraphs. 
The \emph{maximum} $b$-matching problem in hypergraphs is NP-hard and hard to approximate~\cite{elouali2012hyperbmatch,parekh2016genhypermatch}, whereas a
maximal $b$-matching can be computed in linear time by \seqGreedy. Our
goal is to compute a maximal $b$-matching in hypergraphs in parallel.

We write $r = \max_{e \in E} |V(e)|$ for the \emph{rank} of $G$, its
maximum edge size, and $\Delta$ for its maximum degree, the largest number of
edges that contain one vertex. We measure the size of $G$ by its number of vertex--edge incidences,
$m' = \sum_{e \in E} |V(e)|$. We call each incidence a \emph{pin}, i.e., a pair
$(v, e)$ with $v \in V(e)$, thus $m'$ is the number of pins in~$G$. For ordinary graphs
$m' = 2m$. We assume that every vertex lies on at least one edge, thus $n \leq m'$.
Since edges are nonempty, we have $m \leq m'$. 

A \emph{priority order} on the set of edges is a bijection $\pi : E \to \{1, \dots, m\}$. 
We say that edge~$e$ has \emph{higher priority} than $e'$ if $\pi(e) < \pi(e')$. 
A \emph{random order} on edges is one chosen uniformly at random among all $m!$ bijections.
The \emph{sequential greedy} algorithm (\seqGreedy, \Cref{alg:seqgreedy}) considers the edges from the highest to the lowest priority
and takes each edge whose all vertices have some capacity available. The matching
it returns is the \emph{lexicographically first} maximal $b$-matching with respect
to~$\pi$. Our proposed parallel algorithm computes this same matching for a given~$\pi$
(\Cref{th:batch-lexfirst}). We take~$\pi$ to be a
random order, the assumption under which \seqGreedy can be parallelized
efficiently~\cite{BFS}.

Throughout the paper, an event holds \emph{with high probability} (w.h.p.) if it fails
with probability at most $m^{-c}$, for a constant~$c$ that can be made
arbitrarily large. The constants hidden in the bounds may depend on $c$. Since
$m \leq m'$, a failure probability of at \hbox{most $(m')^{-c}$ is also enough.}

\subsection{The fork-join cost model. }

We analyze our algorithms in the fork-join model with work and depth as the
measures of cost, the same model used in~\cite{BB}.
A computation starts as a single task. Any task
may \emph{fork} child tasks, which execute in parallel and asynchronously. The
parent resumes when all of its children have finished (the \emph{join}), and
forks may be nested. Two operations are \emph{dependent} if one precedes the
other in the series--parallel order induced by forks and joins. The operations
of a task are ordered by program order, a task's operations before a fork
precede those of the forked children, and the children's operations precede the
operations the parent performs after the corresponding join. Otherwise they are
concurrent. The \emph{work} of a computation is the total number of
operations it performs, and its \emph{depth} (or span) is the length of its
longest chain of dependent operations. We adopt the binary-forking convention,
under which a fork spawns exactly two
children~\cite{blelloch2020binaryforking}. A \code{parfor} over $k$ items then
expands into a balanced spawn tree, which adds $\mathcal{O}(k)$ work and $\mathcal{O}(\log k)$
depth beyond the cost of the iterations. Blelloch and Brady~\cite{BB} allow a task to
fork any number of children at once. Binary forking is a special case, so our
bounds hold unchanged in their model.

Tasks share memory, and concurrent reads and writes are permitted. As
in~\cite{BB}, the values read must be consistent with some sequential ordering
of the operations that respects the dependences above, so concurrent writes to
the same location have a well-defined but nondeterministic outcome. 
Bounds in this model transfer to the PRAM, a computation with work~$w$ and depth~$d$ can
be simulated on a $p$-processor PRAM in $\mathcal{O}(w/p + d\log^{*} w)$ time
w.h.p.~\cite{BB}.

\subsection{Sequential greedy maximal $b$-matching algorithm (\seqGreedy). }\label{sec:seqgreedy}

\begin{algorithm}[b!]
\small\raggedright
\caption{\seqGreedy: sequential greedy for maximal $b$-matching}
\label{alg:seqgreedy}
\begin{algorithmic}[1]
\INPUT{$G=(V,E)$: a hypergraph; $b:V\to\mathbb{Z}_{\geq 1}$: capacities; $\pi$: a priority order on $E$}
\OUTPUT{$M$: a maximal $b$-matching}
\Statex
\State $M \gets \emptyset$
\State $\bcount(v) \gets 0$ for all $v \in V$
\For{$i = 1, \dots, m$}
  \State let $e$ be the $i$-th edge in $\pi$
  \If{$\bcount(v) < b(v)$ for all $v \in V(e)$}
    \State $M \gets M \cup \{e\}$
    \State $\bcount(v) \gets \bcount(v) + 1$ for all $v \in V(e)$
  \EndIf
\EndFor
\State \Return $M$
\end{algorithmic}
\end{algorithm}

The sequential greedy algorithm for computing a maximal $b$-matching in
hypergraphs, called \seqGreedy\ and shown in \Cref{alg:seqgreedy}, processes
edges according to a priority order $\pi$. It maintains a matching~$M$ and a
counter $\bcount(v)$ for each vertex $v$, holding the number of edges of $M$
incident on~$v$. An edge $e$ is added to $M$ if and only if every endpoint of
$e$ still has capacity available ($\bcount(v) < b(v)$ for all $v \in V(e)$),
and $\bcount(v)$ is then increased by one at each endpoint of $e$. We call
$b(v) - \bcount(v)$ the \emph{remaining capacity} of~$v$, and $v$ is
\emph{saturated} when it reaches zero. The output is the lexicographically
first maximal $b$-matching with respect to $\pi$. The algorithm runs in
$\mathcal{O}(m')$ time. %
Testing an edge $e$ and incrementing the
counters at its endpoints take $\mathcal{O}(|V(e)|)$, which is $\mathcal{O}(m')$ summed over all
edges, and $n \leq m'$ and $m \leq m'$ (\Cref{sec:prelim}). Maximality follows from the
fact that any edge not in $M$ was rejected at an endpoint saturated by
previously added edges, and $\bcount$ never decreases.

\section{Parallel batch-vote maximal $b$-matching algorithm (\parallelBatchBMatch). }\label{sec:batch-vote}

We first introduce some of the notation used in describing the algorithm and state some assumptions.
Every vertex $v$ stores its incident edges in $\edges(v)$ in the priority
order induced by $\pi$. $\Nv(v)$ is the set of \emph{alive} edges incident on~$v$.
Initially $\Nv(v)$ holds all edges of $\edges(v)$, and each round of the algorithm removes from these
sets the edges that are not alive, thus membership in $\Nv(v)$ coincides with
being alive. An edge that is no longer alive is \emph{dead}, and the flag
$\done(e)$ records this. We write $\Nv(U) = \bigcup_{v \in U} \Nv(v)$. 
Throughout, $M$ denotes the matching built so far, and $\deg(v) = |\edges(v)|$, so that
$\sum_{v \in V} \deg(v) = \sum_{e \in E} |V(e)| = m'$.

Blelloch and Brady~\cite{BB} parallelize \seqGreedy\ for unit capacities by letting
every vertex \emph{vote}. A vertex points at its highest-priority alive edge, denoted
by~$\topp(v)$, and votes for it by incrementing an atomic counter $\counter(e)$, which
thus counts the vertices of~$e$ that have voted for it. Once every vertex of~$e$ has voted,
that is, once $\counter(e) = |V(e)|$, the alive edge $e$ becomes a \emph{root} and
joins the matching. A root is the highest-priority alive edge at each of its vertices,
so \seqGreedy\ selects it as well, and two roots never share a vertex, so all roots
of a round may be added at once. A round then removes the roots together with
every other edge incident on a newly matched vertex.

A possible direct route to obtain a parallel maximal $b$-matching algorithm from 
this scheme is to keep its voting
rule intact and change only what a vertex remembers, i.e., give each vertex the same
single pointer $\topp(v)$, have it count the matched edges it has collected, and
remove an edge not when one of its vertices is matched but when one of them
reaches its capacity. This generalized version 
is correct, but it makes
every vertex a \emph{single-vote} vertex, and that is \hbox{expensive for two reasons. }

First, a vertex holding one pointer votes for one edge per round, thus it
acquires at most one match per round and needs $b(v)$ rounds merely to fill its
slots, and $\beta = \max_{v \in V} b(v)$ rounds for the largest capacity. 
For example on a star graph whose center has
capacity $\beta$ and leaves have capacity one, every leaf votes for its only
edge at initialization and then waits. The center's single pointer then lets
exactly one edge hold both of its votes in any round, and execution takes $\beta$
rounds for \emph{every} priority order. Second, 
 rounds are spent on one vertex refilling a pointer it could have filled all at
once.

\begin{algorithm}[t]
\small\raggedright
\caption{\parallelBatchBMatch: parallel batch-voting maximal $b$-matching}
\label{alg:pbbm}
\begin{algorithmic}[1]
\INPUT{$G=(V,E)$: a hypergraph; $b:V\to\mathbb{Z}_{\geq 1}$: capacities}
\OUTPUT{$M$: a maximal $b$-matching}
\Statex
\State $\pi \gets \getRandomPermutation()$
\ParFor{$v \in V$}
  \State $\edges(v) \gets \sortop(\{e \mid v \in e\},\ \pi)$
  \State $\bcount(v) \gets 0$
  \LongState{$\vote(v) \gets$ $b(v)$ highest-priority
             edges of $\edges(v)$}
  \LongState{$\nextp(v) \gets$ the first entry of $\edges(v)$ after entries in $\vote(v)$}
  \ParFor{$e \in \vote(v)$} 
   \State $\increment(\counter(e))$ 
  \EndParFor
\EndParFor
\Statex
\State $W \gets \{e \in E \mid \counter(e) = |V(e)|\}$
\State $M \gets \emptyset$
\While{$|W| > 0$}
  \State $M \gets M \cup W$ %
  \ParFor{$v \in V(W)$}
    \State $\bcount(v) \gets \bcount(v) + |\{e \in W : v \in e\}|$
  \EndParFor
  \Statex
  \State $\Sat \gets \{v \in V(W) \mid \bcount(v) = b(v)\}$
  \State $\finished \gets W \cup \Nv(\Sat)$
  \ParFor{$e \in \finished,\ v \in V(e)$}
    \State $\done(e) \gets \true$
    \State $\deleteop(\Nv(v),\ e)$;\quad $\deleteop(\vote(v),\ e)$
  \EndParFor
  \Statex
  \ParFor{$v \in V(\finished) \setminus \Sat$}
    \LongState{refill $\vote(v)$: advance $\nextp(v)$ along $\edges(v)$, adding
               each alive edge, until $|\vote(v)| = b(v) - \bcount(v)$ or
               $\edges(v)$ is exhausted}
    \ParFor{$e$ newly added to $\vote(v)$}
      \State $\increment(\counter(e))$ %
    \EndParFor
  \EndParFor
  \LongState{$A \gets$ edges newly added to a $\vote$ set this round}
  \State $W \gets \{e \in A \mid \counter(e) = |V(e)|\}$ %
\EndWhile
\State \Return $M$
\end{algorithmic}
\end{algorithm}

Our proposed parallel \emph{batch-vote} maximal $b$-matching algorithm (\parallelBatchBMatch) 
removes the above serialization, a vertex with remaining capacity
$b(v) - \bcount(v)$ votes for its top $b(v) - \bcount(v)$ alive edges at once, where $\bcount(v)$ is the number of edges incident to~$v$ that belongs to the current matching. Thus, it can
acquire many matches in a single round. On the star graph example considered above, it finishes in one round.
Here, $\vote(v)$ is the set of edges for which vertex $v$ currently votes. As before,
$\counter(e)$ counts the number of votes for~$e$ and $e$ is a \emph{root} when
$\counter(e) = |V(e)|$.

\parallelBatchBMatch, shown in Algorithm~\ref{alg:pbbm}, differs from the
unit-capacity scheme of Blelloch and Brady~\cite{BB} in two aspects: (i) it tracks
$\bcount(v)$, and (ii) it replaces the single pointer $\topp(v)$ (i.e., single
vote) by a set of votes. Each vertex~$v$ maintains
the set $\vote(v)$ of its top $b(v) - \bcount(v)$ alive edges, or of all alive
edges if fewer remain, together with a pointer $\nextp(v)$ which indicates the first entry
in the ordered set $\edges(v)$ after the entries that have been added to $\vote(v)$. 
The computation begins by drawing a random priority $\pi$, sorting each vertex's
incident edges by $\pi$, initializing $\bcount$, and having every vertex cast its votes 
(Lines~1--8). The edges
that already gather a full count form the first \emph{root set}~$W$ (Line~9). 

After a round commits the roots $W$, every vertex updates in parallel its $\bcount$ (Lines~13--14). 
The newly saturated vertices are placed in $\Sat$, the set of saturated vertices.
The saturated vertices and their incident edges are removed, with each removed edge deleted
from the $\Nv$ and $\vote$ sets of its vertices (Lines~15--19). 
Each surviving affected vertex then refills $\vote(v)$ to its new remaining capacity by advancing
$\nextp(v)$, incrementing the counters of the edges it newly votes for (Lines~20--23). The
next root set is collected from~$A$, the set of edges newly voted for in the
round, where an edge gaining votes from several endpoints at once is collected
once (Lines~24--25). This loses no root, an alive
edge whose counter was already full when the previous root set was formed
belonged to that root set and was committed and removed, so a new root must
have gained a vote in the current round. 
Unlike under single voting, the root
set $W$ may contain several edges incident to the same
vertex, since a vertex now casts several votes. %

In \parallelBatchBMatch, a vertex votes for an edge $e$ by incrementing the edge's counter, $\counter(e)$, thus
several vertices may increment the same counter within a round. The pseudocode
treats \increment\ as a primitive; we charge each call $\mathcal{O}(1)$ work, and the
calls of a round, $\mathcal{O}(\log m')$ depth in total. Blelloch and Brady~\cite{BB} realize it
in the model by grouping a round's
increments by counter (a semisort) and applying each group's sum as a single
update, in expected amortized work linear in the number of increments and
logarithmic depth w.h.p. A hardware fetch-and-add provides the
operation directly. The two realizations differ in when an increment becomes
visible, and \parallelBatchBMatch\ is not affected by the difference. It defers every
counter test past the round's join (\Cref{alg:pbbm}), thus either realization
serves it. The same grouping primitive builds the sets
of a round, removing duplicates at expected work linear in the number of items.
We import two further primitives from~\cite{BB}. The first is \findNext, which
advances an index past entries failing a predicate, at $\mathcal{O}(d)$ work and
$\mathcal{O}(\log d)$ depth for a slide of distance~$d$. The second is the deletion
from a set at $\mathcal{O}(1)$ amortized work. \parallelBatchBMatch\ also needs the extension of
\findNext\ that returns the next~$k$ entries satisfying the predicate; doubling
to a window that contains $k$ of them and selecting by prefix sums costs
$\mathcal{O}(d)$ work and $\mathcal{O}(\log^2 d)$ depth.
In Appendix \ref{sec:example}, we provide an example illustrating the execution of 
\parallelBatchBMatch\ on a small hypergraph.

\subsection{Correctness. }

We first show that the votes are persistent and identify the invariants of the while loop of \parallelBatchBMatch, and then, 
show that the algorithm terminates and returns a lexicographically \hbox{first maximal $b$-matching.}

\begin{lemma}[Vote persistence]\label{lem:batch-persistence}
In every run of \parallelBatchBMatch, votes are \emph{persistent}, that is, each vertex
increments $\counter(e)$ at most once, counters never decrease, and $\vote(v)$
loses an edge only when that edge is removed. Consequently, while $e$ is alive,
$\counter(e)$ equals the number of vertices currently voting for $e$.
\end{lemma}

\begin{proof}
The proof is given in Appendix~\ref{appendix:proofs}.
\end{proof}

\begin{lemma}[Loop invariants]\label{lem:batch-invariants}
At the start of every iteration of the \textup{while} loop in \parallelBatchBMatch, and when the loop
terminates, the following invariants hold:
\begin{itemize}
  
  \item[\textup{(I1)}] every alive edge has all of its vertices unsaturated.
  \item[\textup{(I2)}] for every unsaturated vertex $v$, the set $\vote(v)$
        consists of the $b(v) - \bcount(v)$ highest-priority alive edges incident
        on $v$, or of all alive edges incident on $v$ if fewer remain.
  \item[\textup{(I3)}] the current root set $W$ consists of exactly the alive
        edges with a full counter.

  \item[\textup{(I4)}] $M$ is a valid $b$-matching. Indeed $\bcount(v)$ is the
        number of edges of $M$ incident on $v$, and $\bcount(v) \leq b(v)$ for
        every $v \in V$.
\end{itemize}
\end{lemma}

\begin{proof}
The proof is given in Appendix~\ref{appendix:proofs}.
\end{proof}

\begin{theorem}[Correctness]\label{thm:batch-correct}
\parallelBatchBMatch\ terminates 
and, at termination, every edge is removed, and the returned~$M$ is a maximal
$b$-matching.
\end{theorem}

\begin{proof}
Every iteration of the while loop in \parallelBatchBMatch\ commits and removes 
at least one edge, and therefore, the while loop terminates. 
The highest-priority alive edge sits in every endpoint's vote set, thus, 
its counter is full and (I3) puts it into the root set. And an unmatched edge 
is removed only at a saturated vertex, \hbox{which stays saturated.}

Each iteration commits its nonempty root set and removes it as part of
$\finished$. Root sets hold only alive edges (I3), and a removed edge is never
alive again, thus each iteration consumes at least one fresh edge, and the while loop 
terminates after at most $m$ iterations.

Next, no edge is alive when the loop exits. Suppose otherwise, and let
$e^\star$ be the highest-priority alive edge. By (I1) every endpoint of
$e^\star$ is unsaturated, and $e^\star$ is its highest-priority alive edge,
thus (I2) places $e^\star$ in every endpoint's vote set, and by persistence of votes (Lemma~\ref{lem:batch-persistence}),
$\counter(e^\star) = |V(e^\star)|$. By (I3) the current root set contains
$e^\star$, therefore it is nonempty and the loop cannot have exited.

Finally, $M$ is maximal. An edge is removed only as a member of
$\finished = W \cup \Nv(\Sat)$, and a round's~$W$ is added to~$M$ before the
marking, thus a removed edge outside $M$ entered $\Nv(\Sat)$, and some vertex
$v$ of it had $\bcount(v) = b(v)$ in that round. Since $\bcount$ never
decreases and never exceeds $b(v)$ (I4), at exit~$v$ still lies in~$b(v)$
edges of~$M$, and adding the edge would exceed its capacity. Every edge is
matched or removed unmatched, therefore~$M$ is maximal.
\end{proof}

The key correctness property is that batching does not change \emph{which}
matching is produced. This requires care: unlike under single voting, a vertex
can acquire a lower-priority match while a higher-priority incident edge is still
alive, when that edge has yet to receive the votes of its other vertices. Thus, the
matches at a vertex need not arrive in priority order. The proof instead uses three
consequences of Lemma \ref{lem:batch-invariants}. First, a vertex never votes for an
edge without also voting for every alive higher-priority incident edge, since by
(I2) the vote set is a priority prefix of the alive edges. Second, a vote, once
cast, is kept while its edge is alive. Third, a vertex saturates only in a round that commits every edge of its vote set (Lemma~\ref{lem:saturation}, below). The last consequence rules out the
bad case, a vertex $v$ saturating while a higher-priority incident edge
$e$ is still alive. Any vote $v$ casts for a lower-priority edge places $e$ in
$\vote(v)$ as well, and $e$ then stays there as long as it lives. The round in
which $v$ saturates commits all of $\vote(v)$, hence $e$ itself, rather than
removing it. 
The proof uses the following fact about saturating vertices.

\begin{lemma}[Saturation]\label{lem:saturation}
If a vertex $u$ saturates in some round, then that round commits every edge of
$\vote(u)$ as it is at the start of the round, and $\vote(u)$ is not empty.
Also, the returned matching $M$ has exactly $b(u)$ edges at $u$, and each of
them has higher priority than every edge at $u$ that is alive at the start of
that round and not in $\vote(u)$.
\end{lemma}

\begin{proof}
Let $C_u$ be the set of roots at $u$ committed in the round. It is not empty,
since $\bcount(u)$ grows in the round. Each edge of $C_u$ is alive with a full
counter (I3), so $u$ votes for it by \Cref{lem:batch-persistence}. Thus
$C_u \subseteq \vote(u)$. The vertex $u$ lies on the alive edges of $C_u$, so
it is unsaturated at the start of the round (I1). Since $u$ saturates,
$\bcount(u) + |C_u| = b(u)$. By (I2), $|\vote(u)| \leq b(u) - \bcount(u) = |C_u|$,
so $C_u = \vote(u)$.

Now let $f$ be an edge at $u$ that is alive at the start of the round and not
in $\vote(u)$. Since~$f$ is alive and not in $\vote(u)$, (I2) says that
$\vote(u)$ is the priority prefix of the alive edges at $u$ of size
$b(u) - \bcount(u)$. Thus, every edge of $C_u = \vote(u)$ has higher priority
than~$f$. Next, let $h$ be an edge committed at $u$ in an earlier round. At
the start of that round, $h$ was in $\vote(u)$, $u$ was unsaturated since it
lies on the alive edge $h$ (I1), and $f$ was alive as well. If $h$ had lower
priority than $f$, then by (I2) $f$ would have been in $\vote(u)$ with it, and
by \Cref{lem:batch-persistence} it would have stayed there. Thus, $h$ has higher
priority than $f$. Finally, the edges of $M$ at $u$ are the $\bcount(u)$ edges
committed at $u$ in earlier rounds (I4) and the edges of $C_u$, since $u$ has
no alive edge after it saturates (I1) and roots are alive (I3). Their count
$\bcount(u) + |C_u| = b(u)$.
\end{proof}

\begin{theorem}[Lexicographically first
maximal $b$-matching]\label{th:batch-lexfirst}
For every priority $\pi$, \parallelBatchBMatch\ returns the lexicographically first
maximal $b$-matching \hbox{with respect to $\pi$}. %
\end{theorem}

\begin{proof}
One direction is a counting argument: a matched edge occupies a
slot at each endpoint, so each endpoint holds fewer than $b(v)$
higher-priority matches, and by induction \seqGreedy sees the same counts and
takes the edge. The other direction rules out the one way batching could lose
an edge $e$ taken by \seqGreedy, an endpoint saturating while $e$ is alive. The saturating
round commits the entire vote set, thus, either $e$ is in it and is committed
too, or the vote set and the earlier matches at that endpoint form $b(v)$
higher-priority matched edges. In the latter case, \seqGreedy filled the
endpoint before reaching~$e$.

Let $M_g$ be the matching obtained by \seqGreedy under $\pi$. 
\seqGreedy adds an edge $e$ iff every endpoint $v$ is incident to fewer than $b(v)$
higher-priority edges of $M_g$, since it takes the edges in priority order and
never removes one, thus, the edges it holds when it reaches $e$ are exactly the
higher-priority edges of $M_g$. We prove that $e \in M \Leftrightarrow e \in M_g$
by induction on $\pi(e)$, assuming the equivalence for all
higher-priority edges.

$(\Rightarrow)$ Suppose $e \in M$. At each endpoint $v$, the edges of $M$
incident on $v$ number at most $b(v)$ (I4), and $e$ is one of them, thus, at
most $b(v) - 1$ of them have higher priority than $e$. By induction, the
higher-priority edges of $M$ at $v$ are exactly the higher-priority edges of
$M_g$ at~$v$, therefore every endpoint of $e$ holds fewer than $b(v)$
higher-priority edges of $M_g$, and \seqGreedy \hbox{adds $e$: $e \in M_g$.}

$(\Leftarrow)$ Suppose $e \in M_g$, thus, each endpoint $v$ is incident to
$c_v < b(v)$ higher-priority edges of $M_g$, which by induction are exactly
the higher-priority edges of $M$ incident on $v$. Assume that no round
commits $e$. Every edge is removed when the loop exits
(\Cref{thm:batch-correct}), and an uncommitted edge is removed only through
$\Nv(\Sat)$. Thus, some round, call it $R$, has an endpoint $v \in V(e)$ in its
$\Sat$ with $e$ alive at its start. If $e \in \vote(v)$ at that start, round $R$ commits
all of $\vote(v)$ (Lemma~\ref{lem:saturation}), hence $e$, a contradiction. Thus, $e \notin \vote(v)$.
By Lemma~\ref{lem:saturation}, $v$ carries $b(v)$ matched edges of higher priority than $e$.
By induction, all of them lie in $M_g$, thus $c_v \geq b(v)$, contradicting
$c_v < b(v)$. Hence, some round commits $e$: $e \in M$.
\end{proof}

In particular, \parallelBatchBMatch\ produces the same output no matter how the
parallel work within \hbox{each round is scheduled.}

\subsection{Work analysis. }

Batching changes how many votes a vertex casts at once, not how often it touches
any one pin: by Lemma~\ref{lem:batch-persistence} a vertex votes for a given edge
at most once, and $\vote(v)$ is refilled by scanning $\edges(v)$ forward, so every
cost can be charged to the pins. Deleting a dead edge from a vote set charges its
pin once more, and collecting the root set from $A$ costs $\mathcal{O}(1)$ per
vote cast in the round, \hbox{at most $m'$ in total.}

\begin{theorem}[Work]\label{thm:batch-work}
\parallelBatchBMatch\ performs $\mathcal{O}(m')$ work in expectation over the
random priority order and the randomization of the primitives of
\Cref{sec:batch-vote}.
\end{theorem}

\begin{proof}
Charge everything to pins. The scans that fill $\vote(v)$ cost
$\mathcal{O}(\deg(v))$ in total, and a pin $(v,e)$ is voted once
and deleted once from $\Nv(v)$ and once from $\vote(v)$. The root set of a
round is collected among the round's new votes, so it, too, is paid for by
pins. The permutation, the sort, and the batched \hbox{increments are
expected-linear.}

The randomized costs are unchanged. Generating $\pi$ takes $\mathcal{O}(m)$
work, bucket sorting the lists $\edges(v)$ by random priorities takes
$\mathcal{O}(m')$ work in expectation (as in the proof
of~\cite[Theorem~3.2]{BB}), and the batched realization of \increment\ costs
expected amortized $\mathcal{O}(1)$ work per call (see above). The
sets a round builds ($V(W)$, $\Sat$, $\Nv(\Sat)$, $\finished$,
$V(\finished) \setminus \Sat$, and $A$) are realized by the same grouping
primitive, at expected work linear in the pins they touch.
The scans that fill $\vote(v)$, at initialization and in the
refills, only advance $\nextp(v)$, and one that advances it by $d$ costs
$\mathcal{O}(d)$, thus, the scans cost $\mathcal{O}(m')$ over all vertices. Every
other pin operation occurs a constant number of times over the whole run.
Also $v$ increments $\counter(e)$
at most once (Lemma~\ref{lem:batch-persistence}). When $e$ is removed, the
deletion pass processes the pin once, marking $e$ and deleting it from
$\Nv(v)$ and $\vote(v)$, since each edge enters $\finished$ at most once,
being removed, and struck from every $\Nv$ set, in its round. The
$b$-matching bookkeeping is also $\mathcal{O}(1)$ per pin. Each vertex enters
$\Sat$ at most once: a saturated vertex has no alive incident edge (I1), and
root sets hold only alive edges (I3), so it never re-enters $V(W)$. Hence
the $\Sat$ filter and the $\bcount$ updates run over the pins of the round's
matched edges, and the construction of $\Nv(\Sat)$, $\finished$, and
$V(\finished) \setminus \Sat$ runs over the pins of the round's finished
edges. Every edge is matched at most once and finished at most once, so all
of this is $\mathcal{O}(1)$ per pin. A round collects its root set among its
newly voted edges $A$, at $\mathcal{O}(1)$ per new vote, and each pin is
voted at most once. The initialization, the scan forming the initial root
set, and the loop control cost
$\mathcal{O}(m' + m + n) = \mathcal{O}(m')$ in total, using $n \leq m'$ and
the bound of at most $m$ iterations (\Cref{thm:batch-correct}). Summing the
per-pin charges and adding the randomized costs gives
$\mathcal{O}(m')$ expected work.
\end{proof}

\subsection{Fork--join structure. }

\parallelBatchBMatch\ is a sequence of rounds, and each round is a nested fork-join
computation: the outer \code{parfor} forks one thread per vertex, and each vertex
thread itself forks one thread per vote to increment edge counters.
The \texttt{while} loop is serial: each round's fork
tree must fully join before the next round begins, so the rounds form a chain of
join barriers. The depth of a round is that of its spawn trees,
$\mathcal{O}(\log m')$, plus the cost of the heaviest leaf, which is a vertex
refilling its vote set. With the $k$-entry form of \findNext\ (introduced above) that refill costs $\mathcal{O}(\log^2 m')$ depth, so a round
has depth $\mathcal{O}(\log^2 m')$. Taken literally as a serial scan the refill
would instead cost $\Omega(\deg(v))$ in a single round, so this primitive is not
a convenience but a requirement. The depth is the number of rounds times the
per-round depth. Bounding the number of rounds is the subject of
\Cref{sec:analysis}.

\subsection{Depth analysis. }\label{sec:analysis}

The fork--join structure of \Cref{sec:batch-vote} leaves two quantities to
bound: the depth of one round and the number of rounds. 

\begin{lemma}[Depth of a round]\label{lem:round-depth}
Each round of \parallelBatchBMatch has $\mathcal{O}(\log^2 m')$ depth w.h.p.
\end{lemma}

\begin{proof}
A round is a constant number of \code{parfor}s whose bodies are primitive
calls, so its depth is the depth of the costliest primitive plus the spawn
trees. Each \code{parfor} ranges over at most $m'$ pins and adds
$\mathcal{O}(\log m')$ depth. Committing $W$ and marking \done\ have
$\mathcal{O}(1)$ bodies. The deletion pass is a batch of at most $m'$ deletions
from the sets $\Nv(\cdot)$ and $\vote(\cdot)$, grouped by set and applied at
$\mathcal{O}(\log m')$ depth w.h.p.~\cite{BB}. The \bcount\ updates and the \increment\ calls
of a round are grouped sums at $\mathcal{O}(\log m')$ depth, and the sets of a
round ($\Sat$, $\finished$, $A$, and the new root set) are built by the same
grouping primitive at logarithmic depth w.h.p.\ 
(primitive introduced when describing the algorithm in \Cref{sec:batch-vote}. A primitive
of~\cite{BB} on an input of size $k \leq m'$ has $\mathcal{O}(\log k)$
depth w.h.p.\ in $k$. As in~\cite{BB}, we take the failure probability
relative to $m'$ rather than $k$, which gives $\mathcal{O}(\log m')$ depth with failure
probability at most $(m')^{-c}$ for the constant $c$ of \Cref{sec:prelim},
whatever $k$ is. The one
step above $\mathcal{O}(\log m')$ is the refill: the extension of \findNext\
that returns the next alive entries up to the remaining capacity costs $\mathcal{O}(\log^2 d)$ depth
for a slide of distance $d \leq \deg(v)$. The refills run in parallel, so a
round has \hbox{depth $\mathcal{O}(\log^2 m')$ w.h.p.}
\end{proof}

Next, we bound the number of rounds and the depth of the algorithm.

\begin{theorem}[\parallelBatchBMatch depth]\label{thm:depth}
For every priority order, \parallelBatchBMatch performs at most $m$ rounds.
Its depth is $\mathcal{O}(m \log^2 m')$ w.h.p.
\end{theorem}

\begin{proof}
Every round commits its root set. The root set is not empty, since the
highest-priority alive edge is always a root (proof of \Cref{thm:batch-correct}).
A root is alive by (I3) of \Cref{lem:batch-invariants}, the round removes it as
part of $\finished$, and a removed edge is never alive again. Thus, every round
commits at least one edge that no earlier round committed, and there are at
most~$m$ rounds. By \Cref{lem:round-depth}, applied with the exponent raised by
one, every round has $\mathcal{O}(\log^2 m')$ depth w.h.p. A union bound over
the at most $m \leq m'$ rounds keeps the failure probability within the
convention of \Cref{sec:prelim}. The initialization generates $\pi$ and sorts
the lists $\edges(v)$ in $\mathcal{O}(\log m')$ depth w.h.p.~\cite{BB}, and its
scans of the lists cost no more depth than a refill. Multiplying the number of
rounds by the depth of a round gives the bound.
\end{proof}

The bound of \Cref{thm:depth} is loose. It holds for every priority order, and
it does not depend on the rank~$r$, on the capacities of vertices, or on the degrees of the
vertices. On the star graph example of \Cref{sec:batch-vote}, the algorithm
finishes in one round for every priority order, while the bound allows~$m$
rounds. Under a uniformly random priority order the algorithm may perform far
fewer rounds, but the current bound does not capture this. Proving a tighter bound on
the number of rounds in terms of~$r$ and the largest vertex capacity~$\beta$ is left
to future work.

\section{Experimental evaluation. }\label{sec:experiments}
In this section, we report our experimental results for the proposed algorithm and competitors.
We run experiments on uniform  capacity $b(v)\in\{1,3,5\}$. All prior parallel competitors are limited to capacity $b\equiv 1$. 
The priorities of our matching are set to negative edge size. This prioritizes smaller edges first and should lead to bigger sized matchings.
We use the edge id as tie-breaker. Furthermore, we test the impact of perturbing the priorities with random numbers from the start.
\subsection{Methodology. }
Our experiments were run on a machine outfitted with a NVIDIA RTX 4090 and a 16-core Intel Xeon w5-3435X running at 3.10 GHz equipped with a L3 cache of 45 MB and 128 GB RAM.
We repeated every experiment three times, while ordering all runs randomly and never running experiments in parallel.
We use the median for averaging runs of one instance and the geometric mean to average across instances.
The running time for algorithms on the GPU does not include the conversion and copying of the hypergraph to it, since our algorithm is usually a step in a bigger computation. The CPU competitor algorithms' running time also do not include the initial setup of the data structures and reads from disk.
\paragraph{Data set. }
 We use the data set described by Reinstädtler~et~al.~\cite{reinstadtler2026hypergraph} for hypergraph matching. The~91 instances stem from the \emph{SAT competition 2014}~\cite{sat2014}, non-linear programming problems~\cite{schenk2009inertia}, general matrices from the \emph{Suite Sparse Collection}~\cite{davis2011}, the \emph{DAC challenge}~\cite{viswanathan2012} and data collected from wikipedia~\cite{reinstadtler_2025_15778413}.
\paragraph{Compared algorithms. }
Main competitor is the algorithm \competitorHLMCUDA from~\cite{reinstadtler2026hypergraph} that uses small random noise to speed up computation and limited to capacity $b\equiv 1$.
The \textsf{Greedy} method in~\cite{grossmann2026hyperbmatch} uses data structures optimized to run a local search afterward, so we provide a simpler and faster implementation called \competitorGREEDYSLIM (1.95 times on average).
We implemented our two versions of our algorithm~\parallelBatchBMatch with batch and single voting \algoFont{(SV)} in CUDA 12.8.
The sorting of the adjacent edges is done in parallel on the GPU and included in the overall running time. If the number of adjacent edges   exceeds ten thousand, sorting is done on the CPU and transfer time for this is included in the running time.
\subsection{Results. }
We discuss the results for capacity~$1$ matchings and higher capacities~$b\in\{3,5\}$. Furthermore, we show the impact of changed orderings induced by random noise. 
\subsubsection{1-matching. }
\begin{figure}
    \centering
    \includegraphics[width=\linewidth]{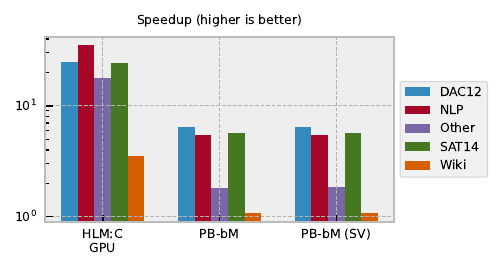}
    \caption{Speedup over \competitorGREEDYSLIM for Hypergraph 1-Matching. \competitorHLMCUDA uses random noise to speed-up computation.}
    \label{fig:exp:1:matching:time}
\end{figure}

For capacity of $b\equiv 1$, we compare with \competitorHLMCUDA, which is a true GPU implementation by~\cite{reinstadtler2026hypergraph} using small random perturbations to speed up their matching process between rounds.
\begin{figure}
    \centering
    \includegraphics[width=0.9\linewidth]{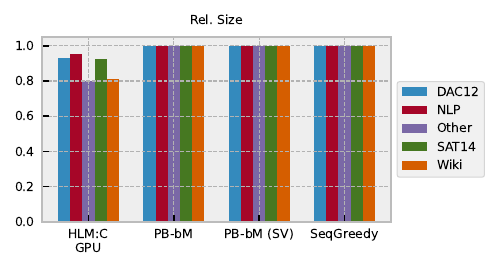}
    \caption{Quality Comparison for Hypergraph 1-Matching.  \parallelBatchBMatch achieves the same results as \competitorGREEDYSLIM.}
    \label{fig:exp:1:matching:quality}
\end{figure}
\paragraph{Quality comparison. } Our algorithms produces the same matchings as the \competitorGREEDY algorithms, as shown in Figure~\ref{fig:exp:1:matching:quality}.
As expected, the voting mechanism does not impact the quality.
The competing \competitorHLMCUDA produces  worse results (reaching 79.9 to 92.6 \% in size) since it uses small perturbations on the edge weights to speed up computations.
\paragraph{Running time. } Figure~\ref{fig:exp:1:matching:time} shows the speedup over the \competitorGREEDYSLIM algorithm. Our algorithms are slower than \competitorHLMCUDA by a factor of 3.28x to 9.7x, but still achieve a speedup of 1.07 to 6.43 times.
Arguably, this difference can be explained by the stricter priority that our algorithm follows and the resulting bigger matchings. There is no difference between single-vote and the batch voting algorithm, as the capacity is $1$ and per vertex we cast one vote at max.
\subsubsection{$b$-matching. } 
\begin{figure}[t]
    \centering
    \includegraphics[width=\linewidth]{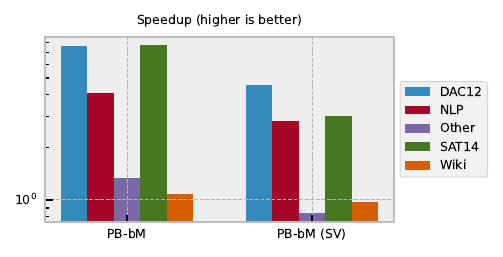}
    \caption{Speedup over \competitorGREEDYSLIM for capacity $b\equiv 3$.}
    \label{fig:exp:3:matching:time}
\end{figure}

\begin{figure}[t]
    \centering
    \includegraphics[width=\linewidth]{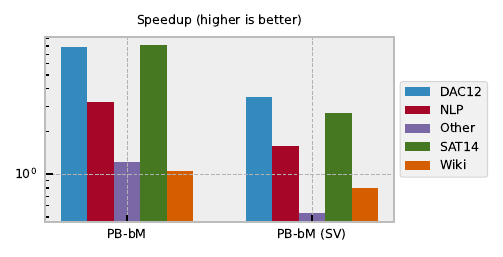}
    \caption{Speedup over \competitorGREEDYSLIM for capacity $b\equiv 5$.}
    \label{fig:exp:5:matching:time}
\end{figure}

For higher capacities of~$3$ and~$5$, our algorithms match the output of state-of-the-art greedy CPU methods.
There is no difference and we place the plots in the Appendix (Figure~\ref{fig:app:exp:3:weight} and~\ref{fig:app:exp:5:weight}).
\paragraph{Running time. }
The speedups over the \competitorGREEDYSLIM algorithm are shown in Figure~\ref{fig:exp:3:matching:time} and~\ref{fig:exp:5:matching:time}.
Our new algorithms are up to 7.68 times faster than the \competitorGREEDYSLIM baseline when using the edge sized induced priorities.
On \dataFont{SAT14} hypergraphs we get the highest speedups, closely followed by the \dataFont{DAC12} instances,  similar to the results on $b\equiv 1$.
The \dataFont{Wiki}  and \dataFont{Other} instances are the most challenging instances in the data set, with  a speedup less than 1.08 and  1.33 times respectively.
This is most-likely due to the structure of the hypergraphs.
The difference between single-vote and batch vote is measurable for $b\equiv 3$, allowing multiple votes enables an average relative speedup of between 1.11 (\dataFont{Wiki}) and 2.55 times (\dataFont{SAT14}). For the \dataFont{Wiki} instance we see a slow down to 0.97 times the running time of the greedy baseline.
For higher capacities ($b\equiv 5$) on the \dataFont{Other} hypergraphs,  single voting slows down to 0.53 times the greedy
  baseline, whereas batch voting maintains a 1.05x speedup, running 1.98x faster.
The overall gap reaches $3$x on the \dataFont{SAT14} instances.


\subsection{Priorities and randomness. } 
We added uniformly distributed noise of two levels ($U=10$ and $U=100$) to randomize the ordering induced by the edge size.
The randomization was done beforehand, simulating changed orderings. We ran the \competitorGREEDYSLIM and our algorithms on these new priorities.

\begin{figure}[t!]
    \centering
    \includegraphics[width=\linewidth]{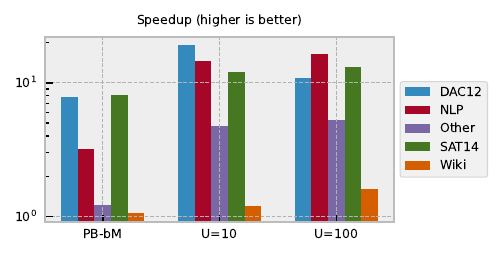}
    \caption{Speedup over \competitorGREEDYSLIM with different levels of uniform randomness with given priority ($U=0,10,100$). Capacity is $b\equiv 5$.}
    \label{fig:abl:5:running}
\end{figure}

\paragraph{Running time. } In Figure~\ref{fig:abl:5:running} the relative speed-up to the \competitorGREEDYSLIM method on the respective level of noise is shown.
For noise of up to size $U=10$ we can see a speed-up for our method to 19.07 times on the \dataFont{DAC12} instances.
Interestingly, with increasing noise ($U=100$) for these hypergraphs the speed-up drops to 10.92.
This trend also holds for all except the \dataFont{Wiki} instance, which gains a higher speed up with more noise.

Overall we can see that our algorithm benefits from more randomized orderings with less tie-breaking by id.

\begin{figure}[t!]
    \centering
    \includegraphics[width=\linewidth]{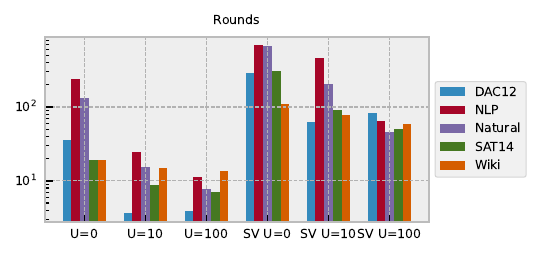}
    \caption{Number of rounds with different levels of uniform randomness ($U=10, 100$) for batch and single vote (SV). Capacity is $b\equiv 5$.}
    \label{fig:abl:5:rounds}
    \end{figure}

\begin{figure}
    \centering
    \includegraphics[width=\linewidth]{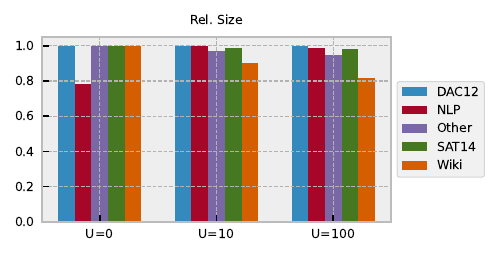}
    \caption{Relative size of matching with different levels of uniform randomness applied to the edge sized induced priority ($U=0,10, 100$). Capacity is $b\equiv 5$.}
    \label{fig:abl:5:size}
\end{figure}

\paragraph{Number of rounds. } Figure~\ref{fig:abl:5:rounds} shows the number of rounds performed with batch and single-voting.
We can clearly see that the number of rounds is reduced with higher randomness in the given permutation when using batch voting.
For single voting, the number of rounds does  stay the same or increase with higher randomness for some data sets.
\paragraph{Quality. } There is no general trend in the relative size of the matchings when applying random noise, see Figure~\ref{fig:abl:5:size}. 
While in most sets, the size decreases (\dataFont{Other},\dataFont{SAT14}) or stays the same (\dataFont{DAC12}), in the \dataFont{NLP} instances it even increases.
However, both algorithms, our and the greedy baseline, produce the same matchings on a given priority.

\begin{figure}[t!]
    \centering
    \includegraphics[width=\linewidth]{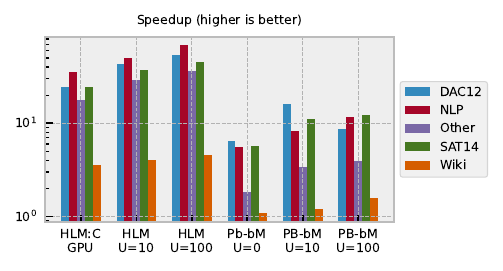}
    \caption{Speedup over \competitorGREEDYSLIM with different levels of uniform randomness in the given priority ($U=0,10,100$). Capacity is $b\equiv 1$.}
    \label{fig:abl:1:running}
\end{figure}

\begin{figure}[t!]
    \centering
    \includegraphics[width=\linewidth]{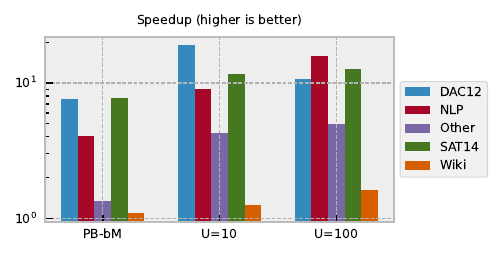}
    \caption{Speedup over \competitorGREEDYSLIM with different levels of uniform randomness in the given priority($U=0,10,100$). Capacity is $b\equiv 3$.}
    \label{fig:abl:3:running}
\end{figure}

We also tested the impact of noise for lower capacities $b\equiv 1,3$ (Figure~\ref{fig:abl:1:running} \&~\ref{fig:abl:3:running}) and were able to reproduce the speed-ups by noise. For $b\equiv 1$ introducing noise also helps the \competitorHLMCUDA algorithm, reaching speed ups of up to 68 times.

In general, we can observe that the priorities have a high impact on the overall performance of all parallel algorithms. The competitor \competitorHLMCUDA relaxes the orderings and produces worse matchings, while our methods are guaranteed to produce the matching induced by the priorities.

\section{Conclusion. }\label{sec:conclusions}

We presented \parallelBatchBMatch, a parallel algorithm for maximal
$b$-matching in hypergraphs. Every vertex votes at once for as many of its
highest-priority edges as it has capacity left. An edge joins the matching once
all of its vertices have voted for it. For a given priority order the algorithm
returns the lexicographically first maximal $b$-matching, i.e., the same
matching as the sequential greedy algorithm. The output does not depend on how
the rounds are scheduled or on the number of cores. The algorithm performs
$\mathcal{O}(m')$ expected work in the binary fork--join model, where $m'$ is
the number of pins.
The algorithm performs at most $m$ rounds and has depth
$\mathcal{O}(m \log^2 m')$ w.h.p. This bound is loose, since it does not depend
on the rank or on the capacities. Proving a tighter bound on the number of
rounds under a random priority order is left to future work.

Batching is what removes the serialization that vertex capacities introduce. A
vertex that votes for one edge per round fills its slots one at a time. Our
experiments show this effect. For capacity $b(v) = 3$, batch voting is on
average between $1.12$ and $2.54$ times faster than single voting, and it needs
fewer rounds once the priorities carry more randomness.
With higher level of randomness in the  priorities, our algorithms can produce a speedup of up to 19.07 times over the greedy baseline, while producing the same matching.

\vfill \pagebreak
\bibliographystyle{siamplain}
\bibliography{references}

\begin{thebibliography}{10}

\bibitem{ahmed2017diverse}
{\sc F.~Ahmed, J.~P. Dickerson, and M.~Fuge}, {\em Diverse weighted bipartite
  $b$-matching}, in Proceedings of the 26th International Joint Conference on
  Artificial Intelligence (IJCAI), 2017, pp.~35--41,
  \url{https://doi.org/10.24963/ijcai.2017/6}.

\bibitem{ahn2014near}
{\sc K.~J. Ahn and S.~Guha}, {\em Near linear time approximation schemes for
  uncapacitated and capacitated $b$-matching problems in nonbipartite graphs},
  in Proceedings of the 25th Annual ACM-SIAM Symposium on Discrete Algorithms
  (SODA), SIAM, 2014, pp.~239--258,
  \url{https://doi.org/10.1137/1.9781611973402.18}.

\bibitem{albers2025stochastic}
{\sc S.~Albers and S.~Schubert}, {\em Online {$b$}-matching with stochastic
  rewards}, in Proceedings of the 50th International Conference on Current
  Trends in Theory and Practice of Computer Science (SOFSEM), vol.~15538 of
  Lecture Notes in Computer Science, Springer, 2025, pp.~37--50,
  \url{https://doi.org/10.1007/978-3-031-82670-2_4}.

\bibitem{alon1986fast}
{\sc N.~Alon, L.~Babai, and A.~Itai}, {\em A fast and simple randomized
  parallel algorithm for the maximal independent set problem}, Journal of
  Algorithms, 7 (1986), pp.~567--583,
  \url{https://doi.org/10.1016/0196-6774(86)90019-2}.

\bibitem{anstee1987polynomial}
{\sc R.~P. Anstee}, {\em A polynomial algorithm for $b$-matchings: An
  alternative approach}, Information Processing Letters, 24 (1987),
  pp.~153--157, \url{https://doi.org/10.1016/0020-0190(87)90178-5}.

\bibitem{bayati2011belief}
{\sc M.~Bayati, C.~Borgs, J.~Chayes, and R.~Zecchina}, {\em Belief propagation
  for weighted $b$-matchings on arbitrary graphs and its relation to linear
  programs with integer solutions}, SIAM Journal on Discrete Mathematics, 25
  (2011), pp.~989--1011, \url{https://doi.org/10.1137/090753115}.

\bibitem{beame1990parallel}
{\sc P.~Beame and M.~Luby}, {\em Parallel search for maximal independence given
  minimal dependence}, in Proceedings of the 1st Annual ACM-SIAM Symposium on
  Discrete Algorithms (SODA), SIAM, 1990, pp.~212--218.

\bibitem{behnezhad2023exponentially}
{\sc S.~Behnezhad, M.~Hajiaghayi, and D.~G. Harris}, {\em Exponentially faster
  massively parallel maximal matching}, Journal of the ACM, 70 (2023),
  pp.~34:1--34:18, \url{https://doi.org/10.1145/3617360}.
\newblock Preliminary version in FOCS 2019.

\bibitem{sat2014}
{\sc A.~Belov, D.~Diepold, M.~Heule, and M.~Järvisalo}, {\em The {SAT}
  competition 2014}.
\newblock \url{http://www.satcompetition.org/2014/index.shtml}, 2014.

\bibitem{bhattacharya2017improved}
{\sc S.~Bhattacharya, M.~Gupta, and D.~Mohan}, {\em Improved algorithm for
  dynamic {$b$}-matching}, in Proceedings of the 25th Annual European Symposium
  on Algorithms (ESA), vol.~87 of Leibniz International Proceedings in
  Informatics (LIPIcs), Schloss Dagstuhl -- Leibniz-Zentrum f{\"u}r Informatik,
  2017, pp.~15:1--15:13, \url{https://doi.org/10.4230/LIPIcs.ESA.2017.15}.

\bibitem{bhattacharya2015dynamic}
{\sc S.~Bhattacharya, M.~Henzinger, and G.~F. Italiano}, {\em Design of dynamic
  algorithms via primal-dual method}, in Proceedings of the 42nd International
  Colloquium on Automata, Languages, and Programming (ICALP), vol.~9134 of
  Lecture Notes in Computer Science, Springer, 2015, pp.~206--218,
  \url{https://doi.org/10.1007/978-3-662-47672-7_17}.

\bibitem{bienkowski2020online}
{\sc M.~Bienkowski, D.~Fuchssteiner, J.~Marcinkowski, and S.~Schmid}, {\em
  Online dynamic {B}-matching: With applications to reconfigurable datacenter
  networks}, ACM SIGMETRICS Performance Evaluation Review, 48 (2020),
  pp.~99--108, \url{https://doi.org/10.1145/3453953.3453976}.

\bibitem{blelloch2026batchmis}
{\sc G.~E. Blelloch, A.~Brady, L.~Dhulipala, J.~T. Fineman, and J.~Lo}, {\em
  Parallel batch-dynamic maximal independent set}, 2026,
  \url{https://arxiv.org/abs/2604.07515}.

\bibitem{BB}
{\sc G.~E. Blelloch and A.~C. Brady}, {\em Parallel batch-dynamic maximal
  matching with constant work per update}, in Proceedings of the 37th ACM
  Symposium on Parallelism in Algorithms and Architectures (SPAA '25), 2025,
  pp.~429--442, \url{https://doi.org/10.1145/3694906.3743313}.

\bibitem{blelloch2020binaryforking}
{\sc G.~E. Blelloch, J.~T. Fineman, Y.~Gu, and Y.~Sun}, {\em Optimal parallel
  algorithms in the binary-forking model}, in Proceedings of the 32nd ACM
  Symposium on Parallelism in Algorithms and Architectures (SPAA '20), 2020,
  pp.~89--102, \url{https://doi.org/10.1145/3350755.3400227}.

\bibitem{BFS}
{\sc G.~E. Blelloch, J.~T. Fineman, and J.~Shun}, {\em Greedy sequential
  maximal independent set and matching are parallel on average}, in Proceedings
  of the 24th ACM Symposium on Parallelism in Algorithms and Architectures
  (SPAA '12), 2012, pp.~308--317,
  \url{https://doi.org/10.1145/2312005.2312058}.

\bibitem{blelloch2011linearwork}
{\sc G.~E. Blelloch, R.~Peng, and K.~Tangwongsan}, {\em Linear-work greedy
  parallel approximate set cover and variants}, in Proceedings of the 23rd ACM
  Symposium on Parallelism in Algorithms and Architectures (SPAA), ACM, 2011,
  pp.~23--32, \url{https://doi.org/10.1145/1989493.1989497}.

\bibitem{brandttumescheit2025dynbsuitor}
{\sc F.~Brandt-Tumescheit, F.~Gerharz, and H.~Meyerhenke}, {\em A fully-dynamic
  approximation algorithm for maximum weight {$b$}-matchings in graphs}, in
  Complex Networks \& Their Applications XIII (COMPLEX NETWORKS 2024),
  vol.~1190 of Studies in Computational Intelligence, Springer, 2025,
  pp.~139--151, \url{https://doi.org/10.1007/978-3-031-82439-5_12}.

\bibitem{buchbinder2007adauctions}
{\sc N.~Buchbinder, K.~Jain, and J.~Naor}, {\em Online primal-dual algorithms
  for maximizing ad-auctions revenue}, in Proceedings of the 15th Annual
  European Symposium on Algorithms (ESA), vol.~4698 of Lecture Notes in
  Computer Science, Springer, 2007, pp.~253--264,
  \url{https://doi.org/10.1007/978-3-540-75520-3_24}.

\bibitem{calkin1990probabilistic}
{\sc N.~Calkin and A.~Frieze}, {\em Probabilistic analysis of a parallel
  algorithm for finding maximal independent sets}, Random Structures \&
  Algorithms, 1 (1990), pp.~39--50,
  \url{https://doi.org/10.1002/rsa.3240010104}.

\bibitem{choromanski2013anonymity}
{\sc K.~Choromanski, T.~Jebara, and K.~Tang}, {\em Adaptive anonymity via
  $b$-matching}, in Advances in Neural Information Processing Systems 26
  (NeurIPS), 2013, pp.~3192--3200.

\bibitem{cook1985taxonomy}
{\sc S.~A. Cook}, {\em A taxonomy of problems with fast parallel algorithms},
  Information and Control, 64 (1985), pp.~2--22,
  \url{https://doi.org/10.1016/S0019-9958(85)80041-3}.

\bibitem{cook2014comparator}
{\sc S.~A. Cook, Y.~Filmus, and D.~T.~M. L{\^e}}, {\em The complexity of the
  comparator circuit value problem}, ACM Transactions on Computation Theory, 6
  (2014), pp.~15:1--15:44, \url{https://doi.org/10.1145/2635822}.

\bibitem{coppersmith1987parallel}
{\sc D.~Coppersmith, P.~Raghavan, and M.~Tompa}, {\em Parallel graph algorithms
  that are efficient on average}, in 28th Annual Symposium on Foundations of
  Computer Science (FOCS), 1987, pp.~260--269,
  \url{https://doi.org/10.1109/SFCS.1987.34}.

\bibitem{cunningham1978primal}
{\sc W.~H. Cunningham and I.~Alfred B.~Marsh}, {\em A primal algorithm for
  optimum matching}, in Polyhedral Combinatorics, vol.~8 of Mathematical
  Programming Studies, Springer, 1978, pp.~50--72,
  \url{https://doi.org/10.1007/BFb0121194}.

\bibitem{davis2011}
{\sc T.~A. Davis and Y.~Hu}, {\em The university of florida sparse matrix
  collection}, ACM Trans. on Math. Software,  (2011),
  \url{https://doi.org/10.1145/2049662.2049663}.

\bibitem{dhulipala2024parallel}
{\sc L.~Dhulipala, M.~Dinitz, J.~{\L}{\k{a}}cki, and S.~Mitrovi\'{c}}, {\em
  Parallel set cover and hypergraph matching via uniform random sampling}, in
  38th International Symposium on Distributed Computing (DISC), vol.~319 of
  LIPIcs, Schloss Dagstuhl -- Leibniz-Zentrum f{\"u}r Informatik, 2024,
  pp.~19:1--19:23, \url{https://doi.org/10.4230/LIPIcs.DISC.2024.19}.

\bibitem{edmonds1973matching}
{\sc J.~Edmonds and E.~L. Johnson}, {\em Matching, {E}uler tours and the
  {C}hinese postman}, Mathematical Programming, 5 (1973), pp.~88--124,
  \url{https://doi.org/10.1007/BF01580113}.

\bibitem{ferdous2021submodularbmatch}
{\sc S.~M. Ferdous, A.~Pothen, A.~Khan, A.~Panyala, and M.~Halappanavar}, {\em
  A parallel approximation algorithm for maximizing submodular {$b$}-matching},
  in Proceedings of the 2021 SIAM Conference on Applied and Computational
  Discrete Algorithms (ACDA), SIAM, 2021, pp.~45--56,
  \url{https://doi.org/10.1137/1.9781611976830.5}.

\bibitem{FN}
{\sc M.~Fischer and A.~Noever}, {\em Tight analysis of parallel randomized
  greedy {MIS}}, ACM Transactions on Algorithms, 16 (2020), pp.~6:1--6:13,
  \url{https://doi.org/10.1145/3326165}.

\bibitem{gabow1983efficient}
{\sc H.~N. Gabow}, {\em An efficient reduction technique for degree-constrained
  subgraph and bidirected network flow problems}, in Proceedings of the 15th
  Annual ACM Symposium on Theory of Computing (STOC), ACM, 1983, pp.~448--456,
  \url{https://doi.org/10.1145/800061.808776}.

\bibitem{gabow2018data}
{\sc H.~N. Gabow}, {\em Data structures for weighted matching and extensions to
  $b$-matching and $f$-factors}, ACM Transactions on Algorithms, 14 (2018),
  pp.~39:1--39:80, \url{https://doi.org/10.1145/3183369}.

\bibitem{gabow2013algebraic}
{\sc H.~N. Gabow and P.~Sankowski}, {\em Algebraic algorithms for $b$-matching,
  shortest undirected paths, and $f$-factors}, in 54th Annual IEEE Symposium on
  Foundations of Computer Science (FOCS), IEEE, 2013, pp.~137--146,
  \url{https://doi.org/10.1109/FOCS.2013.23}.

\bibitem{garrido1992fmatchings}
{\sc O.~Garrido, S.~Jarominek, A.~Lingas, and W.~Rytter}, {\em A simple
  randomized parallel algorithm for maximal $f$-matchings}, in LATIN '92: 1st
  Latin American Symposium on Theoretical Informatics, vol.~583 of Lecture
  Notes in Computer Science, Springer, 1992, pp.~165--176,
  \url{https://doi.org/10.1007/BFb0023827}.

\bibitem{ghaffari2025derand}
{\sc M.~Ghaffari and C.~Grunau}, {\em Towards true work-efficiency in parallel
  derandomization: {MIS}, maximal matching, and hitting set}, in 2025 IEEE 66th
  Annual Symposium on Foundations of Computer Science (FOCS), IEEE, 2025,
  pp.~2694--2727, \url{https://doi.org/10.1109/FOCS63196.2025.00139}.

\bibitem{ghaffari2022mpcbmatch}
{\sc M.~Ghaffari, C.~Grunau, and S.~Mitrovi\'{c}}, {\em Massively parallel
  algorithms for {$b$}-matching}, in Proceedings of the 34th ACM Symposium on
  Parallelism in Algorithms and Architectures (SPAA), 2022, pp.~35--44,
  \url{https://doi.org/10.1145/3490148.3538589}.

\bibitem{grossmann2026hyperbmatch}
{\sc E.~Gro{\ss}mann, F.~Joos, H.~Reinst\"{a}dtler, and C.~Schulz}, {\em
  Engineering hypergraph {$b$}-matching algorithms}, Journal of Graph
  Algorithms and Applications, 30 (2026), pp.~1--24,
  \url{https://doi.org/10.7155/jgaa.v30i1.3166}.

\bibitem{hanguir2021distributed}
{\sc O.~Hanguir and C.~Stein}, {\em Distributed algorithms for matching in
  hypergraphs}, in Approximation and Online Algorithms (WAOA 2020), vol.~12806
  of Lecture Notes in Computer Science, Springer, 2021, pp.~30--46,
  \url{https://doi.org/10.1007/978-3-030-80879-2_3}.

\bibitem{harris2018hypergraphmis}
{\sc D.~G. Harris}, {\em Derandomized concentration bounds for polynomials, and
  hypergraph maximal independent set}, in Proceedings of the 29th Annual
  ACM-SIAM Symposium on Discrete Algorithms (SODA), SIAM, 2018, pp.~2161--2180,
  \url{https://doi.org/10.1137/1.9781611975031.141}.

\bibitem{harris2020distributed}
{\sc D.~G. Harris}, {\em Distributed local approximation algorithms for maximum
  matching in graphs and hypergraphs}, SIAM Journal on Computing, 49 (2020),
  pp.~711--746, \url{https://doi.org/10.1137/19M1279241}.
\newblock Preliminary version in FOCS 2019.

\bibitem{huang2022maxweight}
{\sc C.-C. Huang and F.~Sellier}, {\em Maximum weight {$b$}-matchings in
  random-order streams}, in Proceedings of the 30th Annual European Symposium
  on Algorithms (ESA), vol.~244 of Leibniz International Proceedings in
  Informatics (LIPIcs), Schloss Dagstuhl -- Leibniz-Zentrum f{\"u}r Informatik,
  2022, pp.~68:1--68:14, \url{https://doi.org/10.4230/LIPIcs.ESA.2022.68}.

\bibitem{israeli1986fast}
{\sc A.~Israeli and A.~Itai}, {\em A fast and simple randomized parallel
  algorithm for maximal matching}, Information Processing Letters, 22 (1986),
  pp.~77--80, \url{https://doi.org/10.1016/0020-0190(86)90144-4}.

\bibitem{jebara2009graph}
{\sc T.~Jebara, J.~Wang, and S.-F. Chang}, {\em Graph construction and
  $b$-matching for semi-supervised learning}, in Proceedings of the 26th
  International Conference on Machine Learning (ICML), ACM, 2009, pp.~441--448,
  \url{https://doi.org/10.1145/1553374.1553432}.

\bibitem{kalyanasundaram2000online}
{\sc B.~Kalyanasundaram and K.~Pruhs}, {\em An optimal deterministic algorithm
  for online {$b$}-matching}, Theoretical Computer Science, 233 (2000),
  pp.~319--325, \url{https://doi.org/10.1016/S0304-3975(99)00140-1}.

\bibitem{karp1985fast}
{\sc R.~M. Karp and A.~Wigderson}, {\em A fast parallel algorithm for the
  maximal independent set problem}, Journal of the ACM, 32 (1985),
  pp.~762--773, \url{https://doi.org/10.1145/4221.4226}.

\bibitem{kelsen1992parallel}
{\sc P.~Kelsen}, {\em On the parallel complexity of computing a maximal
  independent set in a hypergraph}, in Proceedings of the 24th Annual ACM
  Symposium on Theory of Computing (STOC), ACM, 1992, pp.~339--350,
  \url{https://doi.org/10.1145/129712.129745}.

\bibitem{khan2018bedgecover}
{\sc A.~Khan, A.~Pothen, and S.~M. Ferdous}, {\em Parallel algorithms through
  approximation: {$b$}-edge cover}, in 2018 IEEE International Parallel and
  Distributed Processing Symposium (IPDPS), IEEE, 2018, pp.~22--33,
  \url{https://doi.org/10.1109/IPDPS.2018.00013}.

\bibitem{khan2016bsuitorsc}
{\sc A.~Khan, A.~Pothen, M.~M.~A. Patwary, M.~Halappanavar, N.~R. Satish,
  N.~Sundaram, and P.~Dubey}, {\em Designing scalable {$b$}-matching algorithms
  on distributed memory multiprocessors by approximation}, in Proceedings of
  the International Conference for High Performance Computing, Networking,
  Storage and Analysis (SC), IEEE, 2016, pp.~773--783,
  \url{https://doi.org/10.1109/SC.2016.65}.

\bibitem{khan2016bsuitorsisc}
{\sc A.~Khan, A.~Pothen, M.~M.~A. Patwary, N.~R. Satish, N.~Sundaram, F.~Manne,
  M.~Halappanavar, and P.~Dubey}, {\em Efficient approximation algorithms for
  weighted {$b$}-matching}, SIAM Journal on Scientific Computing, 38 (2016),
  pp.~S593--S619, \url{https://doi.org/10.1137/15M1026304}.

\bibitem{koufogiannakis2011distributed}
{\sc C.~Koufogiannakis and N.~E. Young}, {\em Distributed algorithms for
  covering, packing and maximum weighted matching}, Distributed Computing, 24
  (2011), pp.~45--63, \url{https://doi.org/10.1007/s00446-011-0127-7}.

\bibitem{lacki2025allocation}
{\sc J.~{\L}{\k{a}}cki, S.~Mitrovi\'{c}, S.~Ramachandran, and W.-H. Sheu}, {\em
  Faster {MPC} algorithms for approximate allocation in uniformly sparse
  graphs}, in Proceedings of the 37th ACM Symposium on Parallelism in
  Algorithms and Architectures (SPAA '25), ACM, 2025, pp.~339--349,
  \url{https://doi.org/10.1145/3694906.3743319}.

\bibitem{liu2023auction}
{\sc Q.~C. Liu, Y.~Ke, and S.~Khuller}, {\em Scalable auction algorithms for
  bipartite maximum matching problems}, in Approximation, Randomization, and
  Combinatorial Optimization (APPROX/RANDOM), vol.~275 of Leibniz International
  Proceedings in Informatics (LIPIcs), Schloss Dagstuhl -- Leibniz-Zentrum
  f{\"u}r Informatik, 2023, pp.~28:1--28:24,
  \url{https://doi.org/10.4230/LIPIcs.APPROX/RANDOM.2023.28}.

\bibitem{luby1986simple}
{\sc M.~Luby}, {\em A simple parallel algorithm for the maximal independent set
  problem}, SIAM Journal on Computing, 15 (1986), pp.~1036--1053,
  \url{https://doi.org/10.1137/0215074}.

\bibitem{mehta2007adwords}
{\sc A.~Mehta, A.~Saberi, U.~Vazirani, and V.~Vazirani}, {\em Adwords and
  generalized online matching}, Journal of the ACM, 54 (2007), pp.~22:1--22:19,
  \url{https://doi.org/10.1145/1284320.1284321}.

\bibitem{mestre2006kextendible}
{\sc J.~Mestre}, {\em Greedy in approximation algorithms}, in Algorithms -- ESA
  2006, vol.~4168 of Lecture Notes in Computer Science, Springer, 2006,
  pp.~528--539, \url{https://doi.org/10.1007/11841036_48}.

\bibitem{miyano1989lexicographically}
{\sc S.~Miyano}, {\em The lexicographically first maximal subgraph problems:
  {P}-completeness and {NC} algorithms}, Mathematical Systems Theory, 22
  (1989), pp.~47--73, \url{https://doi.org/10.1007/BF02088292}.

\bibitem{mueller2000implementing}
{\sc M.~M{\"u}ller-Hannemann and A.~Schwartz}, {\em Implementing weighted
  $b$-matching algorithms: Insights from a computational study}, ACM Journal of
  Experimental Algorithmics, 5 (2000), p.~8,
  \url{https://doi.org/10.1145/351827.384250}.

\bibitem{naim2018gpubmatch}
{\sc M.~Naim and F.~Manne}, {\em Scalable $b$-matching on {GPUs}}, in 2018 IEEE
  International Parallel and Distributed Processing Symposium Workshops
  (IPDPSW), IEEE, 2018, pp.~637--646,
  \url{https://doi.org/10.1109/IPDPSW.2018.00104}.

\bibitem{elouali2012hyperbmatch}
{\sc M.~E. Ouali and G.~J\"{a}ger}, {\em The {$b$}-matching problem in
  hypergraphs: Hardness and approximability}, in Proceedings of the 6th
  International Conference on Combinatorial Optimization and Applications
  (COCOA), vol.~7402 of Lecture Notes in Computer Science, Springer, 2012,
  pp.~200--211, \url{https://doi.org/10.1007/978-3-642-31770-5_18}.

\bibitem{parekh2016genhypermatch}
{\sc O.~Parekh and D.~Pritchard}, {\em Generalized hypergraph matching via
  iterated packing and local ratio}, in Approximation and Online Algorithms
  (WAOA 2014), vol.~8952 of Lecture Notes in Computer Science, Springer, 2015,
  pp.~207--223, \url{https://doi.org/10.1007/978-3-319-18263-6_18}.

\bibitem{reinstadtler2025semistreaming}
{\sc H.~Reinst{\"a}dtler, S.~M. Ferdous, A.~Pothen, B.~U{\c{c}}ar, and
  C.~Schulz}, {\em Semi-streaming algorithms for hypergraph matching}, in 33rd
  Annual European Symposium on Algorithms (ESA), vol.~351 of LIPIcs, Schloss
  Dagstuhl -- Leibniz-Zentrum f{\"u}r Informatik, 2025, pp.~79:1--79:19,
  \url{https://doi.org/10.4230/LIPIcs.ESA.2025.79}.

\bibitem{reinstadtler2026hypergraph}
{\sc H.~Reinst{\"a}dtler, C.~Schulz, N.~Sitchinava, and F.~Walliser}, {\em
  Efficient parallel algorithms for hypergraph matching}, in Euro-Par 2026:
  Parallel Processing, vol.~16782 of Lecture Notes in Computer Science,
  Springer, 2026, pp.~526--540,
  \url{https://doi.org/10.1007/978-3-032-35251-4_36}.

\bibitem{reinstadtler_2025_15778413}
{\sc H.~Reinstädtler, S.~Ferdous, A.~Pothen, B.~Uçar, and C.~Schulz}, {\em
  Wiki hypergraphs for the paper "semi-streaming algorithms for hypergraph
  matching"}, June 2025, \url{https://doi.org/10.5281/zenodo.15778413},
  \url{https://doi.org/10.5281/zenodo.15778413}.

\bibitem{schenk2009inertia}
{\sc O.~Schenk, A.~W{\"a}chter, and M.~Weiser}, {\em Inertia-revealing
  preconditioning for large-scale nonconvex constrained optimization}, SIAM
  Journal on Scientific Computing, 31 (2009), pp.~939--960.

\bibitem{schrijver2003combinatorial}
{\sc A.~Schrijver}, {\em Combinatorial Optimization: Polyhedra and Efficiency},
  vol.~24 of Algorithms and Combinatorics, Springer, 2003.

\bibitem{shun2020practical}
{\sc J.~Shun}, {\em Practical parallel hypergraph algorithms}, in Proceedings
  of the 25th ACM SIGPLAN Symposium on Principles and Practice of Parallel
  Programming (PPoPP), ACM, 2020, pp.~232--249,
  \url{https://doi.org/10.1145/3332466.3374527}.

\bibitem{viswanathan2012}
{\sc N.~Viswanathan, C.~J. Alpert, C.~C.~N. Sze, Z.~Li, and Y.~Wei}, {\em {DAC}
  2012 routability-driven placement contest and benchmark suite}, in DAC '12,
  P.~Groeneveld, D.~Sciuto, and S.~Hassoun, eds., San Francisco, CA, USA, 2012,
  ACM, pp.~774--782, \url{https://doi.org/10.1145/2228360.2228500}.

\end{thebibliography}
\appendix
\section{Extended related work. }\label{sec:related-full}
This appendix gives the full discussion of related work \hbox{summarized in
\Cref{sec:related}.}

\paragraph{Parallel maximal matching and independent set.}
Computing a maximal matching in graphs, or the closely related maximal independent set
(MIS), in parallel is a classical problem addressed by a long line of
algorithms. Karp and Wigderson~\cite{karp1985fast} gave the
first $\mathsf{NC}$ algorithm for MIS, i.e., one running in polylogarithmic time
on polynomially many processors. Luby~\cite{luby1986simple} and Alon, Babai, and
Itai~\cite{alon1986fast} gave randomized $\mathcal{O}(\log n)$-round algorithms which were 
adapted to maximal matching by Israeli and Itai~\cite{israeli1986fast}. 
The degree-constrained
generalization appeared early: Garrido et al.~\cite{garrido1992fmatchings} gave an
$\mathsf{RNC}$ (randomized $\mathsf{NC}$) algorithm for maximal $f$-matchings, the closest classical relative
of parallel $b$-matching. Subsequent work has refined the round and work
complexity substantially, including $\mathcal{O}(\log\log\Delta)$-round maximal matching,
where $\Delta$ is the maximum degree, in the massively parallel (MPC)
model~\cite{behnezhad2023exponentially} and nearly
work-efficient deterministic algorithms~\cite{ghaffari2025derand}. Each of these
computes \emph{some} maximal solution, which depends on the
schedule. None returns the specific solution a given sequential greedy algorithm
would, which is precisely what reproducibility across machines \hbox{and core counts
requires.}

\paragraph{Greedy matching and dependence depth.}
Reproducing the solution obtained by a sequential greedy algorithm 
amounts to computing the lexicographically first maximal solution. For the lexicographically first maximal independent set this
is P-complete~\cite{cook1985taxonomy,miyano1989lexicographically}, so
lexicographically first problems are in general resistant to parallelism under a
fixed adversarial order. For matching on ordinary graphs the picture is different.
That problem is complete for the class $\mathsf{CC}$ rather than known to be
P-complete~\cite{cook2014comparator}. The standard remedy is to
randomize the order. Coppersmith, Raghavan, and
Tompa~\cite{coppersmith1987parallel} and Calkin and
Frieze~\cite{calkin1990probabilistic} first established that the greedy algorithm
is shallow on random graphs. Furthermore, Blelloch, Fineman, and Shun~\cite{BFS}
proved the stronger
statement that on \emph{any} graph a uniformly random order yields a
\emph{dependence depth} of $\mathcal{O}(\log^2 m)$ w.h.p., 
where the dependence depth is the
number of rounds obtained by repeatedly matching, in parallel, every edge having no
higher-priority conflict. Fischer and Noever~\cite{FN} subsequently tightened this
bound to $\mathcal{O}(\log m)$ w.h.p. Building on this framework, Blelloch and
Brady~\cite{BB} obtained a work-efficient parallel algorithm for maximal matching in hypergraphs
with $\mathcal{O}(m')$ expected work and $\mathcal{O}(\log^2 m)$ depth w.h.p. The parallel 
algorithm returns exactly the solution obtained by the sequential greedy maximal matching algorithm. Their approach is the starting point for the design of our
parallel algorithm for maximal $b$-matching in hypergraphs. The same
lexicographically-first-under-random-order principle has recently been extended to the parallel batch-dynamic maximal independent set
problem~\cite{blelloch2026batchmis}. No analogue of these
round bounds is known once vertex capacities are considered. 
In \Cref{sec:analysis}
we give a simple bound on the number of rounds of our algorithm, and we leave a
tighter bound to future work.

\paragraph{Parallel hypergraph matching.}
Parallel algorithms for hypergraph matching have been developed along several
methodological lines. The local-max approach of Reinst\"adtler et
al.~\cite{reinstadtler2026hypergraph} adds, in each round, all locally maximal
edges under freshly randomized weights, in the spirit of Luby's MIS, and terminates
in $\mathcal{O}(\log m)$ rounds. The same authors also gave semi-streaming
algorithms~\cite{reinstadtler2025semistreaming}. In the distributed and massively
parallel models, Harris~\cite{harris2020distributed} and Hanguir and
Stein~\cite{hanguir2021distributed} propose approximation algorithms, while Dhulipala
et al.~\cite{dhulipala2024parallel} gave PRAM (parallel random-access machine)
and MPC algorithms for hypergraph
matching together with the related set cover and packing problems, continuing the
work-efficient parallel-greedy line of Blelloch, Peng, and
Tangwongsan~\cite{blelloch2011linearwork} and the practical shared-memory framework
of Shun~\cite{shun2020practical}. A natural reduction takes maximal hypergraph
matching to MIS on the line graph, to be solved by a parallel hypergraph-MIS
algorithm~\cite{beame1990parallel,harris2018hypergraphmis,kelsen1992parallel}.
However, the line graph may contain $\Omega(m^2)$ edges, so the reduction is far
from work-efficient. None of these algorithms accounts for vertex capacities, and thus they do not
solve the $b$-matching \hbox{problem in hypergraphs.}

\paragraph{Sequential $b$-matching algorithms.}
$b$-matching in graphs is a thoroughly studied generalization of matching. Its polyhedral
structure and exact polynomial-time solvability were established by Edmonds and
Johnson~\cite{edmonds1973matching}, Cunningham and
Marsh~\cite{cunningham1978primal}, and Anstee~\cite{anstee1987polynomial}, and are
surveyed by Schrijver~\cite{schrijver2003combinatorial}. Exact algorithms range
from Gabow's combinatorial reductions and blossom data
structures~\cite{gabow1983efficient,gabow2018data} to algebraic
matrix-multiplication methods~\cite{gabow2013algebraic}, with message-passing and
LP-based solvers used in practice~\cite{bayati2011belief}. On hypergraphs the
problem is considerably harder, computing a \emph{maximum} $b$-matching is NP-hard
and admits no PTAS~\cite{elouali2012hyperbmatch}. The best known polynomial-time
guarantees are LP-relative approximations~\cite{parekh2016genhypermatch}. Thus,
the practical target is a \emph{maximal} solution, produced in linear
time by \seqGreedy. Greedy and approximation algorithms for $b$-matching are
themselves well studied, including a $\tfrac12$-approximation via $k$-extendible
systems in $\mathcal{O}(\beta m)$ time~\cite{mestre2006kextendible}, an early appearance of
the factor $\beta = \max_{v \in V} b(v)$, near-linear-time approximation
schemes~\cite{ahn2014near}, and careful engineering
studies~\cite{grossmann2026hyperbmatch,mueller2000implementing}. All of these,
however, are sequential.

\paragraph{Parallel $b$-matching algorithms.}
Beyond the early $\mathsf{RNC}$ algorithm of Garrido et
al.~\cite{garrido1992fmatchings} mentioned above, the literature on parallel
$b$-matching offers only approximation guarantees, and it
concerns ordinary graphs almost exclusively. For shared memory, the b-Suitor
family~\cite{khan2016bsuitorsc,khan2016bsuitorsisc} computes
$\tfrac12$-approximations that scale to billions of edges, with a GPU
implementation~\cite{naim2018gpubmatch} and extensions to submodular
$b$-matching~\cite{ferdous2021submodularbmatch} and
$b$-edge-cover~\cite{khan2018bedgecover}. Auction algorithms give
$(1-\varepsilon)$-approximate maximum-cardinality bipartite $b$-matchings in the
work--depth model~\cite{liu2023auction}. In the distributed model, weighted
hypergraph $b$-matching has also been
addressed~\cite{koufogiannakis2011distributed}. In the massively parallel (MPC)
model, Ghaffari, Grunau, and Mitrovi\'{c}~\cite{ghaffari2022mpcbmatch} gave a
$(1+\varepsilon)$-approximation in $\mathcal{O}(\log\log\bar d)$ rounds, where $\bar d$ is
the average degree, and recent work
improves the round complexity of approximate \hbox{allocation on sparse
graphs~\cite{lacki2025allocation}.}

\paragraph{Other models.}
Finally, $b$-matching has been studied extensively in the
dynamic~\cite{bhattacharya2017improved,bhattacharya2015dynamic,brandttumescheit2025dynbsuitor},
streaming~\cite{huang2022maxweight}, and
online~\cite{albers2025stochastic,buchbinder2007adauctions,kalyanasundaram2000online,mehta2007adwords}
settings. These models are orthogonal to the problem considered in this paper and the associated algorithms are not discussed in detail and are not listed in Table~\ref{tab:related}.

\section{Illustrative example. }\label{sec:example}
\begin{figure*}[t]
\centering
\definecolor{hedgeone}{HTML}{0072B2}
\definecolor{hedgetwo}{HTML}{E69F00}
\definecolor{hedgethree}{HTML}{009E73}
\newcommand{\hedgeplain}[4][solid]{%
  \begin{scope}[transparency group, opacity=0.55]
    \draw[#2, line width=31pt, line join=round, line cap=round, #1] #4;
    \filldraw[draw=#3, fill=#3, line width=28pt, line join=round, line cap=round] #4;
  \end{scope}}
\newcommand{\hedgebold}[3]{%
  \begin{scope}[transparency group, opacity=0.62]
    \draw[#1, line width=34pt, line join=round, line cap=round] #3;
    \filldraw[draw=#2, fill=#2, line width=28pt, line join=round, line cap=round] #3;
  \end{scope}}
\begin{tikzpicture}[
  >={Stealth[length=1.8mm]},
  vtx/.style={circle, draw=black, fill=white, minimum size=7mm, inner sep=0pt, font=\normalsize},
  elbl/.style={font=\small, inner sep=1.2pt, align=center},
  vlbl/.style={font=\small, inner sep=1pt},
  plbl/.style={font=\normalsize},
  vote/.style={->, black, semithick},
]
\def\pA{(-0.78,0.46) -- (0,2.00) -- (0.78,0.46) -- cycle}     %
\def\pB{(-0.78,0.46) -- (0,-0.91) -- (-1.95,-1.24) -- cycle}  %
\def\pC{(0.78,0.46) -- (0,-0.91) -- (1.95,-1.24) -- cycle}    %
\node[anchor=center, font=\small] at (-4.30,0.30) {\begin{tabular}{@{}cc@{}}
  \toprule
  $v$   & $b(v)$ \\
  \midrule
  $v_1$ & $2$ \\
  $v_2$ & $1$ \\
  $v_3$ & $2$ \\
  $v_4$ & $1$ \\
  $v_5$ & $1$ \\
  $v_6$ & $1$ \\
  \bottomrule
\end{tabular}};
\begin{scope}
  \hedgeplain{hedgeone!75!black}{hedgeone!34}{\pA}
  \hedgeplain{hedgetwo!75!black}{hedgetwo!34}{\pB}
  \hedgeplain{hedgethree!75!black}{hedgethree!34}{\pC}
  \node[vtx] at (-0.78,0.46)  {$v_1$};
  \node[vtx] at (0,2.00)      {$v_2$};
  \node[vtx] at (0.78,0.46)   {$v_3$};
  \node[vtx] at (0,-0.91)     {$v_4$};
  \node[vtx] at (-1.95,-1.24) {$v_5$};
  \node[vtx] at (1.95,-1.24)  {$v_6$};
  \draw[vote] (-0.55,0.61) -- (-0.21,0.83);
  \draw[vote] (-0.82,0.18) -- (-0.87,-0.22);
  \draw[vote] (0,1.72) -- (0,1.35);
  \draw[vote] (0.55,0.61) -- (0.21,0.83);
  \draw[vote] (0.82,0.18) -- (0.87,-0.22);
  \draw[vote] (-0.26,-0.81) -- (-0.64,-0.67);
  \draw[vote] (-1.72,-1.09) -- (-1.38,-0.87);
  \draw[vote] (1.72,-1.09) -- (1.38,-0.87);
  \node[elbl, text=hedgeone!70!black]   at (0,2.95)      {$\pi(e_1){=}1$};
  \node[elbl, text=hedgetwo!70!black]   at (-1.95,-2.20) {$\pi(e_2){=}2$};
  \node[elbl, text=hedgethree!70!black] at (1.95,-2.20)  {$\pi(e_3){=}3$};
  \node[plbl] at (0,-2.90)    {(a) after initialization};
\end{scope}
\draw[black!30] (3.10,3.20) -- (3.10,-3.10);
\begin{scope}[xshift=7.1cm]
  \hedgebold{hedgeone!90!black}{hedgeone!66}{\pA}
  \hedgebold{hedgetwo!90!black}{hedgetwo!66}{\pB}
  \hedgeplain[dashed]{black!30}{black!7}{\pC}
  \node[vtx] at (-0.78,0.46)  {$v_1$};
  \node[vtx] at (0,2.00)      {$v_2$};
  \node[vtx] at (0.78,0.46)   {$v_3$};
  \node[vtx] at (0,-0.91)     {$v_4$};
  \node[vtx] at (-1.95,-1.24) {$v_5$};
  \node[vtx] at (1.95,-1.24)  {$v_6$};
  \node[elbl, text=hedgeone!70!black]   at (0,2.95)      {$e_1 \in M$};
  \node[elbl, text=hedgetwo!70!black]   at (-1.95,-2.20) {$e_2 \in M$};
  \node[elbl, text=black!50] at (1.95,-2.20)  {$e_3 \in E \setminus M$};
  \node[vlbl] at (-2.45,0.90) {$\bcount(v_1){=}2$};
  \node[vlbl] at (2.45,0.90)  {$\bcount(v_3){=}1$};
  \node[plbl] at (0,-2.90)    {(b) after the round};
\end{scope}
\end{tikzpicture}
\caption{Example: \parallelBatchBMatch\ execution on a small hypergraph with three edges. The table on the left lists the capacity of every vertex.}
\label{fig:batch-round}
\end{figure*}
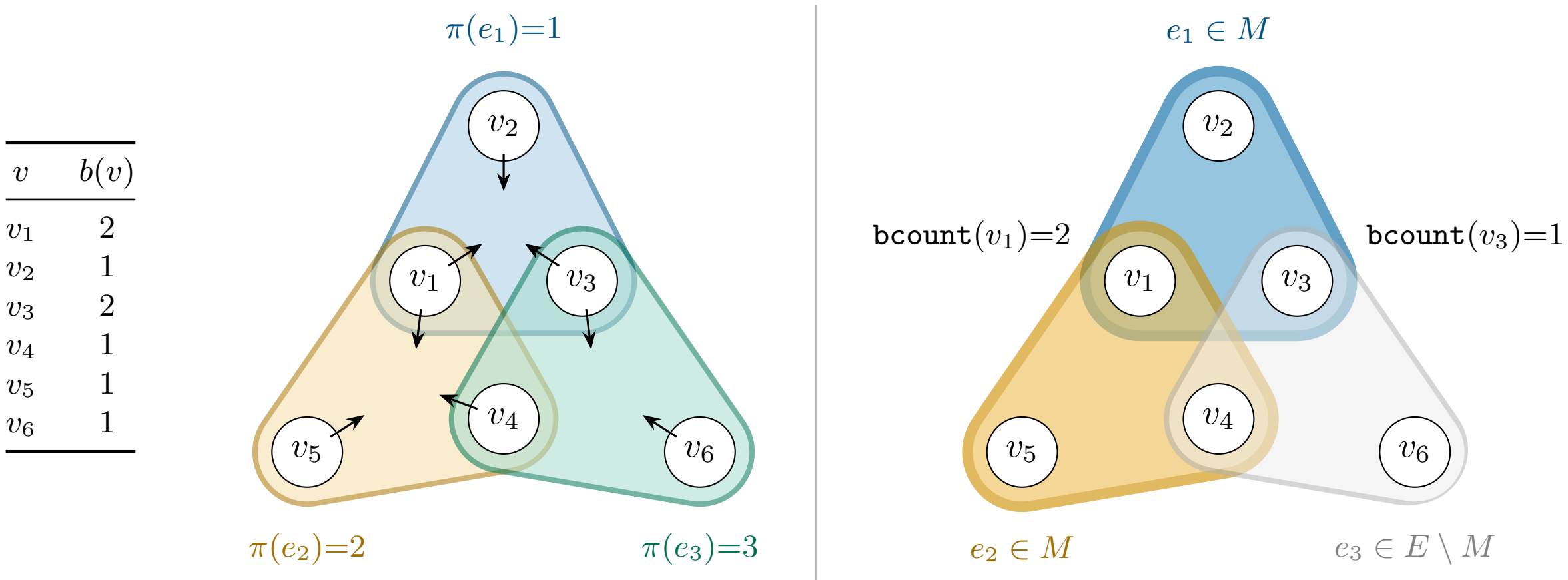

In \Cref{fig:batch-round}, we provide an example of \parallelBatchBMatch\ execution  
on a small hypergraph with three edges. The priority order is 
$\pi(e_1) = 1$, $\pi(e_2) = 2$, $\pi(e_3) = 3$, and the capacity of every
vertex is listed in the table on the left of the figure. Each edge is the shaded region around its vertices, 
in its own color and labeled with its priority $\pi(e)$. In \Cref{fig:batch-round}(b) 
the two matched edges 
are drawn bold and~$e_3$, left unmatched, is in faded gray. 
Arrows indicate the votes cast at initialization. On \Cref{fig:batch-round}(a), vertices~$v_1$ 
and~$v_3$ each have two slots and vote for their top two
edges at once, thus $e_1$ and $e_2$ collect a vote from every endpoint in the same round, while $e_3$
still lacks $v_4$'s vote. The root set $W = \{e_1, e_2\}$ therefore contains two
edges sharing $v_1$, which a single-vote rule could never produce. 
\Cref{fig:batch-round}(b) shows that committing $W$ saturates $v_1, v_2, v_4, v_5$, so $e_3$
is removed through~$v_4$ even though $v_3$ keeps a free slot. The sets $\vote(v_3)$ and
$\vote(v_6)$ are empty, and the execution of \parallelBatchBMatch\ ends after this 
single round. An algorithm allowing $v_1$ only one vote per round would instead need two 
rounds here, one per matched edge.

\needspace{0.3\textheight}
\section{Additional experimental results. }\leavevmode\par
\begin{figure}[H]
    \centering
    \includegraphics[width=\linewidth]{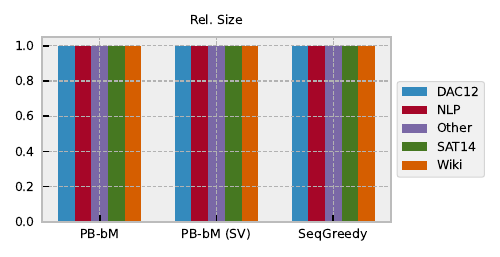}
    \caption{There is no difference in the quality of the computed matchings for $b(v)=3$.}
    \label{fig:app:exp:3:weight}
\end{figure}
\begin{figure}[H]
    \centering
    \includegraphics[width=\linewidth]{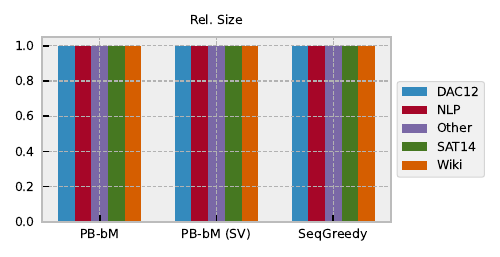}
    \caption{There is no difference in the quality of the computed matchings for $b(v)=5$.}
    \label{fig:app:exp:5:weight}
\end{figure}

\section{Proofs. }\label{appendix:proofs}

\subsection{Proof of Lemma \ref{lem:batch-persistence}. }

\begin{proof}
A vote, once cast, stays until its edge dies, since the deletion
pass is the only place a vote set shrinks, and the scans that fill vote sets
only move forward along $\edges(v)$, thus no edge is voted for twice. Validity
holds because the roots at a vertex all lie in its vote set, which never
exceeds the capacity the vertex has left. No root escapes the collection
from $A$, because an alive edge whose counter was already full at the previous
collection joined that root set and was removed with it.

The set $\vote(v)$ shrinks only in the deletion pass,
which deletes from it exactly the edges of $\finished$, the edges removed in
that round, thus $\vote(v)$ loses an edge only when that edge is removed. The
initialization takes the first $b(v)$ entries of
$\edges(v)$ and leaves $\nextp(v)$ on the entry after them. A refill only
advances $\nextp(v)$, therefore $\nextp(v)$ passes each entry of
$\edges(v)$ at most once, and the entries it has passed form a growing prefix.
An edge enters $\vote(v)$ only when $\nextp(v)$ passes its entry and it is
alive, hence at most once. It cannot return after leaving,
since it leaves only by dying, and the scans add alive edges only. A vertex
increments $\counter(e)$ exactly when it adds $e$ to its vote set, hence at
most once, and only endpoints of $e$ do so, since
$\vote(v) \subseteq \edges(v)$. No counter is ever decremented. Thus,
$\counter(e)$ counts the vertices that have ever added $e$ to their vote
sets. While $e$ is alive, each of them still votes for $e$, thus $\counter(e)$
is the number of current voters, it never exceeds $|V(e)|$, and a full
counter means that every endpoint votes for $e$.
\end{proof}

\subsection{Proof of Lemma~\ref{lem:batch-invariants}. }
\begin{proof}
The invariants hold before the first iteration: $M$ is empty and
$\bcount(v) = 0$. No edge is removed, and $b(v) \geq 1$ leaves every vertex
unsaturated. The set $\vote(v)$ holds the $b(v)$ highest-priority
edges of $\edges(v)$, which is (I2) for $\bcount(v) = 0$. The initial
root set is computed over all of $E$, every edge alive, which is (I3). We
show that one iteration preserves the four, assuming them at its start. The
loop exits at such a start, thus the invariants hold there as well.

(I1). $\bcount$ changes only at vertices of $V(W)$, so a vertex
reaches $b(v)$ in this round exactly if it lies in $V(W)$ with its new count
at $b(v)$, which is membership in $\Sat$. For $v \in \Sat$, the set $\Nv(v)$
holds every edge alive at the round's start that is incident on $v$. All of
them enter $\finished = W \cup \Nv(\Sat)$ and are removed. A vertex saturated
in an earlier round has no alive incident edge already at the round's start
(I1), and $\bcount$ never decreases, thus after the round no alive edge has a
saturated vertex.

(I2). If no edge at $v$ was removed this round, then
$v \notin V(\finished)$, and since the committed edges of any $v \in V(W)$
lie in $W \subseteq \finished$, also $v \notin V(W)$, so the alive edges at
$v$, the count $\bcount(v)$, and $\vote(v)$ are all unchanged, and (I2)
carries over. Otherwise $v \in V(\finished)$. Suppose $v$ is unsaturated when
the next iteration begins. Then $v \notin \Sat$, the vertex was unsaturated
already at the round's start, as $\bcount$ never decreases, and the round
refills $v$.
After the deletion pass, $\vote(v)$ is its old set of edges minus the set of removed edges.
It is still a prefix of the alive edges in priority order: an alive edge of
higher priority than a surviving member was in the old prefix (I2) and,
being alive, was not deleted. It also has room: $\vote(v)$ lost at least its
$|C_v|$ committed members, while $\bcount(v)$ increased by exactly $|C_v|$, so
$|\vote(v)| \leq b(v) - \bcount(v)$ holds for the new count. Every alive edge
among the entries $\nextp(v)$ has passed is already in $\vote(v)$: it was alive
when $\nextp(v)$ passed it, thus it was added, and it has not left since. The refill therefore
extends the prefix by exactly the next alive edges in $\edges(v)$, until
$|\vote(v)| = b(v) - \bcount(v)$ or no entries remain, which restores (I2).

(I3). Every edge of the new root set is collected from $A$ with a full
counter, and it is alive when the next iteration begins: it was alive when
the refill added it, and the round removes nothing after its deletion pass.
Conversely, let $e$ be alive at the next start with $\counter(e) = |V(e)|$.
If $e$ gained no vote in this round, its counter was full already at the
round's start, and~$e$ was alive then, thus $e \in W$ by (I3), and the round
committed and removed it, a contradiction. Therefore, $e$ gained a vote, lies in $A$,
and is collected. Hence the new root set is exactly the set of alive edges
\hbox{with a full counter.}

(I4). Let $v \in V(W)$, and let $C_v = \{e \in W : v \in e\}$ be the
roots committed at $v$. Each $e \in C_v$ is alive with a full counter (I3),
thus, by the persistence of votes (Lemma~\ref{lem:batch-persistence}), every endpoint of $e$ votes for it: $C_v \subseteq \vote(v)$.
The vertex $v$ is unsaturated, its edges in $W$ being alive (I1), thus by (I2)
$|\vote(v)| \leq b(v) - \bcount(v)$, and $\bcount(v)$ increases by
$|C_v| \leq b(v) - \bcount(v)$, to at most $b(v)$. No other $\bcount$
changes. The edges of $W$ are new to $M$: they are alive (I3), while an edge
committed earlier was removed in its round, as a member of its $\finished$,
and stays removed. Thus, $M \cup W$ adds exactly the edges of $W$, and
$\bcount(v)$ remains the number of edges of $M$ incident on $v$.
\end{proof}

\ifblind
\else
\fi

\end{document}